\documentclass[preprint,aps,prb]{revtex4}
\usepackage{amsmath}
\usepackage{amsfonts}
\usepackage{amssymb}
\usepackage{graphicx}
\usepackage{subfigure}
\usepackage{color}
\usepackage{graphicx}
\usepackage{epsfig}

\begin{document}
\title{Optical and Hall conductivities of a thermally disordered two-dimensional spin-density wave: two-particle response in the pseudogap regime of electron-doped high-$T_c$ superconductors }
\author{Jie Lin}
\affiliation{Materials Science Division, Argonne National Laboratory, Argonne, IL 60439}
\author{A. J. Millis}
\affiliation{Department of Physics, Columbia University,
538 West 120th Street, New York, NY 10027}

\begin{abstract}
We calculate the longitudinal ($\sigma_{xx}$)  and Hall ($\sigma_{xy}$) optical conductivities for two-dimensional metals with thermally disordered antiferromagnetism using a generalization of an approximation introduced by Lee, Rice and Anderson for the self energy. The conductivities are calculated from the Kubo formula, with current vertex function treated in a conserving approximation satisfying the Ward identity. In order to obtain a finite DC limit, we introduce phenomenologically impurity scattering, with relaxation time $\tau$. $\sigma_{xx}(\Omega)$ satisfies the $f$-sum rule. For the infinitely peaked spin correlation function, $\chi(\mathbf{q})\propto \delta(\mathbf{q}-\mathbf{Q})$, we recover the expressions for the conductivities in the mean-field theory of the ordered state.  When the spin correlation length $\xi$ is large but finite, both $\sigma_{xx}$ and $\sigma_{xy}$ show behaviors characteristic of the state with long-range order. The calculation runs into difficulty for $\Omega\lesssim 1/\tau$. The difficulties are traced to an inaccurate treatment of the very low energy density of states within the Lee-Rice-Anderson approximation. The results for $\sigma_{xx}(\Omega)$ and $\sigma_{xy}(\Omega)$ are qualitatively consistent with data on electron-doped cuprates when $\Omega>1/\tau$.

\end{abstract}


\maketitle


\section{Introduction}

Long-range antiferromagnetic order can have a profound effect on the electronic excitation spectrum of metals, opening a gap over some or all of the Fermi surface.\cite{Overhauser62}  By continuity, it seems reasonable to believe that even in the absence of long-range order, finite-range correlations may also have an important effect on the electronic excitation spectrum.  The effects may be expected to be particularly large in two-dimensional systems with Heisenberg symmetry,  because in this case long ranged order can only exist at temperature $T=0$. Even in the presence of weak coupling into a third dimension or weak Ising anisotropy, a wide range of temperatures will exist where the physics is controlled by the thermally disordered magnetic state. Such a state, which following the usual conventions we refer to as a spin-density wave (SDW),  is believed to occupy a significant portion of the phase diagram of electron-doped cuprates.\cite{Armitage10} Extensive experimental studies, including optical conductivity,\cite{Zimmers05,Bontemps06,Bontemps07} Hall effect,\cite{Dagan04,Charpentier10} and infrared magnetotransport,\cite{Jenkins09} of these materials in the doping range where the ground state has long-range SDW order, have revealed signatures characteristic of partial gap opening starting at a temperature high compared to the N\'eel temperature $T_N(x)$. This gap seems to be closely related to that in the SDW state, since the measured quantities evolve smoothly across $T_N(x)$.\cite{Charpentier10,Jenkins09} Furthermore, a recent inelastic neutron scattering study\cite{Motoyama07} on the Nd$_{2-x}$Ce$_x$CuO$_{4\pm\delta}$ materials found that the spin correlation length $\xi$ remains large for temperatures high above $T_N$. This motivates the theoretical proposal that the scattering of electrons off thermal spin fluctuations may hold the key to understanding the transport data above $T_N(x)$.\cite{Vilk95,Vilk96}

While the experimental phenomenology is clear, our theoretical understanding of this regime is incomplete. In a seminal paper, Lee, Rice, and Anderson\cite{Lee73} (LRA) proposed a model for the study of  electron dynamics in the presence of long but finite ranged density wave order. In this model, electrons are coupled to quasi-static (relevant frequencies less than $k_BT$) order-parameter fluctuations, resulting in suppression of the single-particle density of states at low energies, a phenomenon sometimes referred to as a ``pseudogap''. The Lee-Rice-Anderson analysis was generalized by Sadovskii\cite{Sadovskii79} and then was extended to two-dimensional systems close to the antiferromagnetic instability by Vilk, Tremblay, and co-workers.\cite{Vilk95,Vilk96}  who argued that such long but finite ranged antiferromagnetic fluctuations controlled important aspects of the physics of  the electron-doped cuprates. In a further theoretical development,  Schmalian \textit{et al} argued that the electron Green's function can be \textit{exactly} calculated for the two-dimensional LRA model of electrons with a cuprate band dispersion scattered from antiferromagnetic spin fluctuations,\cite{Schmalian99} generalizing the method first used by Sadovskii in the study of one-dimensional charge-density fluctuations.\cite{Sadovskii79} Tchernyshyov\cite{Tchernyshyov99} analyzed the underlying assumptions of the Sadovskii's solution, and argued that it should be used with caution in the generical two-dimensional situation. However, he concluded that in particular regions of momentum space termed ``hot spots'',  the method could be safely applied, and it is near these momentum points that Schmalian \textit{et al} found pseudogap behavior in agreement with the previous work of Vilk and Tremblay.\cite{Vilk95,Vilk96} There has been an attempt to calculate the conductivity using this method.\cite{Sadovskii02} However, the restriction to the vicinity of the hot spots makes the Sadovskii solution unsuitable for the study of transport properties  in two-dimensional systems, because a summation over the entire Brillouin zone is needed. A generalization of the LRA theory to transport phenomena is required. 

In this paper, we provide the missing generalization. We use the two-dimensional LRA model in which electrons are coupled to themal (quasi-static) antiferromagnetic spin fluctuations to study the optical and Hall conductivities of electron-doped cuprates at temperatures above $T_N$. As in Refs [\onlinecite{Lee73,Vilk95,Vilk96}], we calculate the electron self-energy in the leading order of perturbation theory. The new feature of our work is a calculation of the current vertex function in a conserving approximation.\cite{Baym61,Baym62} We find that although the vertex function  corresponding to the LRA self energy leads to a conductivity which fulfills the $f$-sum rule,  the  dynamic ($+-$) current vertex function has unphysical features at low frequencies;  leading in some cases to an unphysical negative conductivity in the very low frequency region. The difficulty is traced to an incorrect treatment of the subgap density of states in the LRA calculation. We discuss ways of curing the difficulty and also present results at higher frequencies which are not significantly affected by the problem.

The rest of the paper is organized as follows. In Sec. \ref{model}, we use the spin-fermion model to motivate the LRA model, calculate the electron self-energy in the leading-order perturbation theory, and discuss the pseudogap phenomenon in the resulting single-particle spectral function. In Sec. \ref{conductivity}, we study the optical conductivity with a proper treatment of the current vertex function. In Sec. \ref{Hall}, we study the Hall conductivity, developing a calculation scheme which can reproduce the mean-field result in the proper limit. In Sec. \ref{summary}, we summarize our results, and discuss the implications. Some technical details and a brief summary of the mean-field theory can be found in various Appendices. 


\section{Model and electron self-energy \label{model}}

\begin{figure}[htbp]
\centering
\includegraphics[width=.5\textwidth]{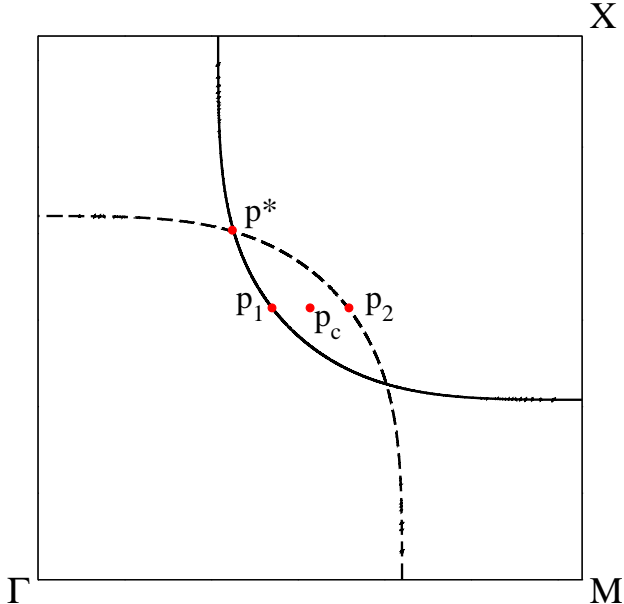}
\caption{\small{Quadrant of two-dimensional Brillouin zone showing the Fermi surface (solid line) for electron-doped cuprates, its translation (`backfolding') by wavevector $\mathbf{Q}=(\pi,\pi)$  (dashed line), `hot spot' $\mathbf{p}^\ast$ and other Fermi surface points referred to in subsequent discussions.}}
\label{fig:fsplots}
\end{figure}

In this section, we present the model, the basic approximation we use, and results for the self energy. The results reproduce those previously derived\cite{Lee73,Sadovskii79,Vilk95,Vilk96,Schmalian99,Tchernyshyov99} and are presented here to establish notation.  Our starting point is electrons moving with a two-dimensional dispersion chosen, for concreteness, to resemble that believed to be relevant to  high-$T_c$ superconductors:\cite{Andersen95} 
\begin{equation}\label{eq:band dispersion}
\varepsilon_p=-2t(\cos p_x+\cos p_y)+4t^\prime\cos p_x \cos p_y-2t^{\prime\prime}(\cos 2p_x+\cos 2p_y)-\mu,
\end{equation}
where $t=0.38$eV, $t^\prime=0.32t$, $t^{\prime\prime}=0.5t^\prime$, and $\mu$ is the chemical potential. Figure \ref{fig:fsplots} shows the resulting Fermi surface for electron doping $x\approx 0.16$ (solid line). Throughout this paper, we choose units such that the lattice constant $a=1$, $\hbar=1$, and measure energy in units of $t$, unless otherwise stated.

We represent the effects of magnetism via the  spin-fermion model, which  has been used extensively in the study of itinerant electrons in systems close to or in long-range magnetically ordered states.\cite{Moriya,Hertz76,Millis93,Ioffe98,Abanov03,Paul05,Lohneysen07} It is a low-energy effective theory with an intrinsic cutoff energy $\Lambda$, and is conveniently formulated as an effective action\cite{Abanov03} 
\begin{equation}
\begin{split}
S=&-\int_0^\beta d\tau\int_0^\beta d\tau^\prime \sum_{p\sigma} c_{p\sigma}^\dagger(\tau)
G_0^{-1}(\mathbf{p},\tau-\tau^{\prime})c_{p\sigma}(\tau^\prime)\\
&+\frac{1}{2}\int_0^\beta d\tau\int_0^\beta d\tau^\prime
\sum_q\chi_0^{-1}(\mathbf{q},\tau-\tau^\prime)\mathbf{S}_q(\tau)\cdot\mathbf{S}_{-q}(\tau^\prime)\\
&+g\int_0^\beta d\tau\sum_{q}\mathbf{S}_{-q}(\tau)\cdot \mathbf{s}_q(\tau),
\end{split}
\label{eq:sf action}
\end{equation}
where $c_{p\alpha}$ is the fermionic field operator, $G_0^{-1}(\mathbf{p},\tau)$ is the inverse of the bare fermionic Green's function, $\mathbf{S}_q$ is an emergent field describing collective antiferromagnetic spin fluctuations which are important to the low-energy physics, $\chi_0(\mathbf{q},\omega)=\chi_0/[\xi^{-2}+(\mathbf{q}-\mathbf{Q})^2-(\omega/v_s)^2]$ is the bare susceptibility in the spin-fermion model with $\mathbf{Q}=(\pi,\pi)$, $\mathbf{s}_q=\sum_p c_{p+q\alpha}^\dagger\boldsymbol{\sigma}_{\alpha\beta}c_{p\beta}$ is the spin density operator of electrons with $\boldsymbol{\sigma}$ the Pauli matrices, and $g$ is the effective coupling constant between electrons and spin fluctuations. 

$G_0(\mathbf{p},\tau)$ is the Fourier transform of 
\begin{equation}\label{eq:G0}
G_0(\mathbf{p},i\epsilon_n)=\frac{1}{i\epsilon_n-\varepsilon_p+\frac{i}{2\tau}\mathrm{sgn}\epsilon_n},
\end{equation}
where $\epsilon_n=(2n+1)\pi T$, and we have explicitly introduced the impurity scattering rate $1/2\tau$, which will be discussed in the next section. Eq (\ref{eq:G0}) has been extensively used in studies of the fluctuation conductivity close to the Peierls transition\cite{Patton73,Takada78} and the superconducting transition.\cite{Larkin,Caroli67}  When two different scattering processes, spin fluctuations and impurities, are present, it is necessary to consider their interference.\cite{Zala01,Paul08} The renormalization of the spin-fermion interaction vertex $g$ by impurity scattering and that of the impurity scattering relaxation time $\tau$ by the spin-fermion interaction are discussed in Appendix \ref{gtau}, where it is demonstrated that both renormalizations are finite. Thus, as long as we keep $g$ and $\tau$ as adjustable parameters of the calculation, we can neglect their mutual renormalizations. 

The third term in Eq (\ref{eq:sf action}) represents the interaction between electrons and spin fluctuations, and effectively arises from an interaction Hamiltonian, 
\begin{equation}
\mathcal{H}_{\mathrm{sf}}=g\sum_{p,q}\mathbf{S}_{-q}\cdot c_{p+q\alpha}^\dagger \boldsymbol{\sigma}_{\alpha\beta}c_{p\beta}.
\label{eq:Hsf}
\end{equation}
In this paper, we consider the state without long-range order, $<\mathbf{S}_q>=0$, and assume that the spin fluctuations are isotropic. As a result, the spin indices on electrons are irrelevant for the calculation of charge transport coefficients. After properly redefining $g$ to account for the three $\mathbf{S}$ directions and two electron spin projections, the interaction Hamiltonian can be written as 
\begin{equation}
H_{\mathrm{sf}}=g\sum_{p,q,\sigma}S_{-q} c_{p+q\sigma}^\dagger c_{p\sigma},
\label{eq:Hsf 2}
\end{equation}
which bears the form of the electron-phonon interaction with $S_q$ playing the role of the phonon field operator $A_q=a_q+a_{-q}^\dagger$.\cite{Mahan} There are important differences between the spin-fermion model in Eq (\ref{eq:Hsf 2}) and the electron-phonon problem. The phonon degrees of freedom are extrinsic to electrons. Because of the large mass of nuclei compared to electrons, the Migdal's theorem applies, which allows the electron-phonon interaction in conventional metals to be treated in a controlled manner.\cite{AGD} However, the spin-fluctuation degrees of freedom in the spin-fermion model are intrinsic to electrons. For the Migdal's theorem to be applicable, one usually resorts to one or another variant of the large-$N$ limit where $N$ is the number of fermion flavors\cite{Altshuler94} or the number of hot spots.\cite{Abanov03} As in the electron-phonon problem, the bare spin-fluctuation propagator $\chi_0(\mathbf{q},\omega)$ is renormalized by creation and annihilation of electron-hole pairs, which leads to the Landau damping term, $i\omega/\omega_{\mathrm{sf}}$. In the random phase approximation, the renormalized spin-fluctuation propagator has the form 
\begin{equation}
\chi(\mathbf{q},\omega)=\frac{\chi_0}{\xi^{-2}+(\mathbf{q}-\mathbf{Q})^2+i\omega/\omega_\mathrm{sf}},
\label{eq:chi}
\end{equation}
where $\omega_\mathrm{sf}/\xi^2$ sets the energy scale for spin fluctuations, and can be expressed as combinations of the parameters in Eq (\ref{eq:sf action}) (see \textit {e.g.} Ref [\onlinecite{Abanov03}]). We note that Eq (\ref{eq:chi}) has the same form as that proposed phenomenologically by Millis \textit{et al},\cite{Millis90} can be obtained from the self-consistent renormalization theory,\cite{Moriya,Moriya00}  and has the generic form in the theory of quantum phase transitions involving itinerant electrons.\cite{Sachdev,Hertz76,Millis93} We thus argue that the applicability of Eq (\ref{eq:chi}) is independent of microscopic details and approximations involved in deriving it. The remaining question is to calculate effects of the interaction in Eq (\ref{eq:Hsf 2}) on fermions. 

When the temperature $T$ is relatively large compared to $\omega_\mathrm{sf}/\xi^2$, the dynamic term in Eq (\ref{eq:chi}) can be neglected (for more discussion, see Appendix \ref{static}, and for a related discussion in the context of superconducting fluctuations, see Ref [\onlinecite{Larkin}]). In terms of Matsubara frequencies, the static spin-fluctuation propagator is written as 
\begin{equation}
\chi(\mathbf{q},i\omega_n)=\frac{\chi_0}{\xi^{-2}+(\mathbf{q}-\mathbf{Q})^2}\delta_{n,0},
\label{eq:chi static}
\end{equation}
which is the two-dimensional generalization of the  LRA model. 

\begin{figure}[htbp]
\centering
\includegraphics[width=.8\textwidth]{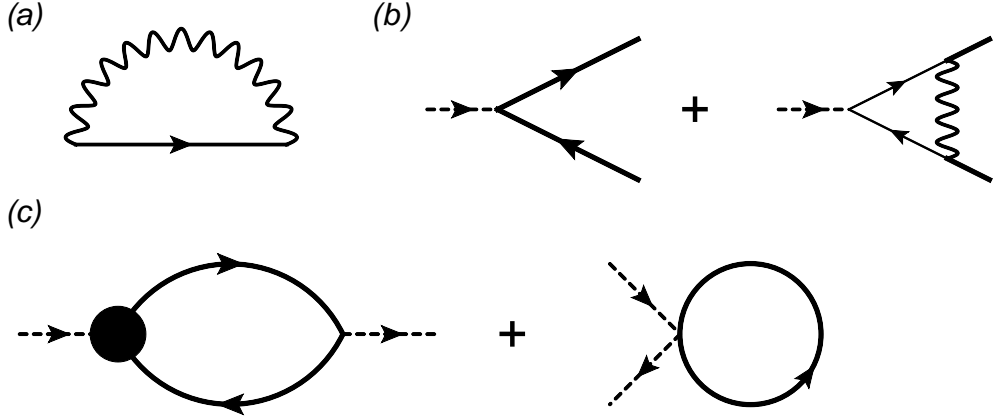}
\caption{\small{Feynman diagrams used in this paper. (a) The Lee-Rice-Anderson approximation to the electron self energy, (b) the current vertex function corresponding to the Lee-Rice-Anderson self energy and (c) diagrams needed for the longitudinal conductivity. The thin solid lines in (a) and (b) represent the bare electron Green's function $G_0$ in Eq (\ref{eq:G0}), the thick solid lines in (b, c) represent the dressed Green's function $G$ in Eq (\ref{eq:G}), the wavy lines are the spin-fluctuation propagator $\chi$ in Eq (\ref{eq:chi static}), and the solid circle in (c) represents the current vertex function $\Gamma^J$ and is calculated according to (b).}}
\label{fig:PT diagrams}
\end{figure}

Using Eq (\ref{eq:G0}) and Eq (\ref{eq:chi static}), we calculate the electron self energy to leading order in $g$, shown in Figure \ref{fig:PT diagrams} (a),
\begin{equation}\label{eq:SE eq}
\Sigma(\mathbf{p},i\epsilon_n)=g^2T\int \frac{d\mathbf{q}}{(2\pi)^2}\frac{\chi_0}{(\mathbf{q}-\mathbf{Q})^2+\xi^{-2}}\frac{1}{i\epsilon_n-\varepsilon_{p+q}+i\mathrm{sgn}\epsilon_n/(2\tau)}. 
\end{equation}
As discussed above, there is a cut-off energy scale $\Lambda$ below which the spin-fermion model is defined. Since the above integral is convergent at large $\mathbf{q}$, we assume that all   energies under consideration are below $\Lambda$. To proceed, we change $\mathbf{q}\to \mathbf{q}+\mathbf{Q}$ and write $\mathbf{q}=(q_{\parallel},q_\perp)$ where $q_\parallel$ and $q_\perp$ are the components parallel and perpendicular to $\mathbf{v}_{p+Q}$, respectively. The $q_\parallel$-integral can be done by the residue method, and the remaining $q_\perp$-integral is elementary. The retarded self-energy is obtained by analytical continuation to the real axis via $i\epsilon_n\to \omega+i\delta$ with $\delta$ a positive infinitesimal, 
\begin{equation}\label{eq:SE}
\Sigma^R(\mathbf{p},\omega)=\frac{\lambda}{i\varepsilon_0}F\Bigl(\frac{\omega-\varepsilon_{p+Q}+i/2\tau}{i\varepsilon_0}\Bigr).
\end{equation}
Here we introduce the effective coupling constant $\lambda$ and energy scale $\varepsilon_0$ defined by
\begin{eqnarray}
\lambda&=& g^2T\chi_0/(2\pi),
\label{lambdadef}\\
\varepsilon_0&=&v_{p+Q}\xi^{-1},
\label{epsilon0def}
\end{eqnarray}
where the weak $p$-dependence of $\varepsilon_0$ will be neglected in subsequent calculations. The function $F$ is given by
\begin{equation}
F(z)=\int_0^\infty\frac{dx}{\sqrt{x^2+1}}\frac{1}{z+\sqrt{x^2+1}}=\frac{1}{\sqrt{z^2-1}}\ln\frac{1+z+\sqrt{z^2-1}}{1+z-\sqrt{z^2-1}}.
\end{equation}

In the limit $1/\tau\to 0$, we reproduce the result of Refs [\onlinecite{Vilk95,Vilk96}], 
\begin{equation}\label{eq:SE if}
\begin{split}
\Sigma_{if}^R(\mathbf{p},\omega)=\frac{\lambda}{\sqrt{(\omega-\varepsilon_{p+Q})^2+\varepsilon_0^2}}\Bigl\{&\mathrm{sgn}(\omega-\varepsilon_{p+Q})\ln \frac{\varepsilon_0}{\sqrt{(\omega-\varepsilon_{p+Q})^2+\varepsilon_0^2}-|\omega-\varepsilon_{p+Q}|}\\
&-i\pi/2\Bigr\}.
\end{split}
\end{equation} 
We note that $\mathrm{Im}\Sigma_{if}^R(\mathbf{p},\omega)<0$ as expected and have verified that $\mathrm{Re}\Sigma_{if}^R(\mathbf{p},\omega)$ and $\mathrm{Im}\Sigma_{if}^R(\mathbf{p},\omega)$ are related by the Kramers-Kr\"onig relation, 
\begin{equation}\label{eq:KK}
\mathrm{Re}\Sigma_{if}^R(\mathbf{p},\omega)=\frac{1}{\pi}\mathcal{P}
\int_{-\infty}^\infty d\bar{\omega}\frac{\mathrm{Im}\Sigma_{if}^R(\mathbf{p},\bar{\omega})}{\bar{\omega}-\omega}.
\end{equation}

Using the one-dimensional analogue of Eq (\ref{eq:SE eq}), one obtains the result of Lee \textit{et al},\cite{Lee73} 
\begin{equation}\label{eq:SE 1d}
\Sigma^R_{1D}(p,\omega)=\frac{\Delta^2}{\omega-\varepsilon_{p+Q}+i/2\tau+i\varepsilon_0},
\end{equation}
where $\Delta^2=g^2T\chi_0\xi/2$. We note that $\Sigma_{1D}^R(p,\omega)$ has a simple pole at $\omega=\varepsilon_{p+Q}-i/2\tau-i\varepsilon_0$ in the lower-half $\omega$-plane, as expected. 

We now discuss the physical content of the results. The retarded Green's function is 
\begin{equation}\label{eq:G}
G^R(\mathbf{p},\omega)=\frac{1}{\omega-\varepsilon_p+i/2\tau-\Sigma^R(\mathbf{p},\omega)},
\end{equation}
and the spectral function $A$ is
\begin{equation}
A(\mathbf{p},\omega)=-2\mathrm{Im}G^R(\mathbf{p},\omega).
\label{Adef}
\end{equation}
For comparison to previous results, we will sometimes present results based on Eq (\ref{eq:SE if}) (\textit{i.e.} for the model without an explicit additional impurity scattering). We denote the corresponding Green's function and spectral function as $G^R_{if}$ and $A_{if}$, respectively. 

The situation is particularly simple in the one-dimensional case. If we linearize the dispersion near the Fermi level $\varepsilon_p\rightarrow vp$, $\varepsilon_{p+Q}\rightarrow -vp$, measure momenta relative to the Fermi momentum $\pm k_F$, and assume $Q=2k_F$, then 
\begin{equation}
G_{1D}^R(p,\omega) =\frac{\omega+vp+\frac{i}{2\tau}+i\varepsilon_0}{\omega^2-(vp)^2-\Delta^2+ i\left(\frac{1}{2\tau}+\varepsilon_0\right)(\omega-vp)+\frac{i}{2\tau}(\omega+vp)},
\label{G1d}
\end{equation}
exhibiting a gap of size $\Delta^2+(vp)^2$ broadened by the impurity scattering rate and by the finite correlation length (parametrized by $\varepsilon_0$). In obtaining this result, it is crucial to use the bare Green's function in Eq (\ref{eq:SE eq}). Self-consistent one-loop approximations (and related approximations such as the fluctuation-exchange approximation (FLEX)) do not obtain a pseudogap.

In the two-dimensional case of main interest here, the situation is more complicated because the Green's function depends both on position on the Fermi surface and on displacement of the momentum away from it. However, a few general statements can be made. We note that Eq (\ref{eq:SE if}) can be written as $\Sigma^R_{if}(\mathbf{p},\omega)=\varepsilon_0\bigl[(\lambda/i\varepsilon_0^2)F(\omega/i\varepsilon_0-\varepsilon_{p+Q}/i\varepsilon_0)\bigr]$. Thus, the spectral function shows scaling behavior: $\varepsilon_0A_{if}(\mathbf{p},\omega/\varepsilon_0)$ is invariant if energies and frequencies are measured in units of $\varepsilon_0$ at fixed $\mathbf{p}$ and $\lambda/\varepsilon_0^2$.

Precise results can be obtained in the limit that the spin-fluctuation propagator $\chi(\mathbf{q})$ is infinitely peaked at $\mathbf{Q}$ (the Kampf-Schrieffer model\cite{Kampf90})
\begin{equation}\label{eq:chi KS}
\chi_{KS}(\mathbf{q},i\omega_n)=\chi_0\delta_{n,0}\delta(\mathbf{q}-\mathbf{Q}).
\end{equation}
Eq (\ref{eq:SE eq}) gives 
\begin{equation}\label{eq:SE KS}
\Sigma^R_{KS}(\mathbf{p},\omega)=\frac{\Delta_s^2}{\omega-\varepsilon_{p+Q}+i/2\tau}, 
\end{equation}
where $\Delta_s^2=g^2T\chi_0/(2\pi)^2$. Substituting this self-energy into Eq (\ref{eq:G}), one obtains the Green's function in the mean-field theory of the SDW state (the diagonal matrix elements in Eq (\ref{eq:G MF})) without introducing a condensate.\cite{Chubukov97} In the next two sections, we shall extend this conclusion to the optical and Hall conductivities; the mean-field expressions for $\sigma_{xx}$ and $\sigma_{xy}$ can be obtained from Eq (\ref{eq:chi KS}) in the leading-order perturbation theory. 

An important role in subsequent discussions is played by the ``hot spots'', momenta $\mathbf{p}^\ast$ such that both $\mathbf{p}^\ast$ and $\mathbf{p}^\ast+\mathbf{Q}$ are on the Fermi surface (Figure \ref{fig:fsplots}).  At these points, the density of states is most strongly reduced from the non-interacting value. The structure of the spectral function at the hot spots is parameterized by a gap scale $\Delta_{pg}$ and the scaling arguments of the previous paragraph show that  $\Delta_{pg}/\varepsilon_0$ depends only on $\lambda/\varepsilon_0^2$.  In the limit $\lambda/\varepsilon_0^2\gg 1$, $\Delta_{pg}/\varepsilon_0\gg 1$, and is determined by the equation
\begin{equation}\label{eq:PGeq}
\Delta_{pg}=\mathrm{Re}\Sigma^R_{if}(\mathbf{p}^\ast,\Delta_{pg}).
\end{equation}
To leading logarithmic accuracy, we find 
\begin{equation}
\Delta_{pg}\approx \sqrt{\lambda}\left(\ln\sqrt{\frac{4\lambda}{\varepsilon_0^2}}\right)^{1/2}.
\label{Deltaapprox}
\end{equation} 
Thus in the two-dimensional case,  in the limit $\lambda/\varepsilon_0^2\gg 1$, $\Delta_{pg}$ is determined mainly by $\sqrt{\lambda}$ with a (weak) logarithmic dependence on $\varepsilon_0$. This equation should be contrasted to the one-dimensional result $\Delta^2=g^2T\chi_0\xi/2$.  

\begin{figure}
\centering
\includegraphics[width=.8\textwidth]{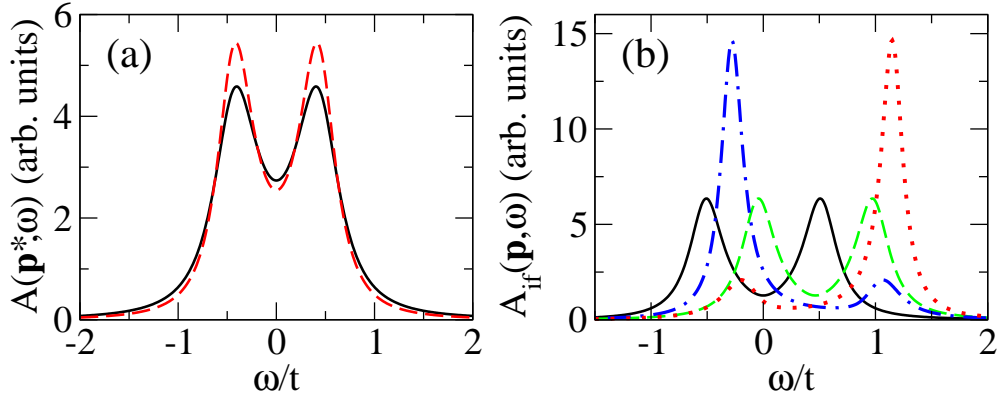}
\caption{\small{(a): comparison of the spectral functions $A(\mathbf{p}^\ast,\omega)$ (solid line) and $A_{if}(\mathbf{p}^\ast,\omega)$ (dashed line) at one hot-spot, $\mathbf{p}^\ast$ (see Fig. \ref{fig:fsplots}). The parameters are $\varepsilon_0=0.2t$, $\lambda=2.5\varepsilon_0^2$, and $1/2\tau=0.05t$. (b): the spectral function $A_{if}(\mathbf{p},\omega)$ at $\mathbf{p}^\ast$ (solid line, black online), $\mathbf{p}_c=(\pi/2,\pi/2)$ (dashed line, green online), $\mathbf{p}_1=(0.43\pi,\pi/2)$ (dash-dotted line, blue online), and $\mathbf{p}_2=(0.57\pi,\pi/2)$ (dotted line, red online). The parameters are $\varepsilon_0=0.1t$ and $\lambda=10\varepsilon_0^2$.}}
\label{fig:Acompare}
\end{figure}

Panel (a) of Figure \ref{fig:Acompare} shows the spectral functions $A(\mathbf{p}^\ast,\omega)$ (including impurity scattering, solid line) and $A_{if}(\mathbf{p}^\ast,\omega)$ (no impurity scattering, dashed line) at the hot spot $\mathbf{p}^\ast$, for $1/2\tau=0.05t$, $\varepsilon_0=0.2t$, and $\lambda=2.5\varepsilon_0^2$. Both curves show suppression of the spectral weight at low frequencies.  We define the pseudogap $\Delta_{pg}$ as half the distance between the two peaks on the corresponding curve and see that the two curves have roughly equal pseudogap values, $\sim 0.4t$, slightly larger than that predicted by the asymptotic result in Eq (\ref{Deltaapprox}), $\approx 0.35t$. This panel thus demonstrates that we can use either Eq (\ref{eq:SE}) or Eq (\ref{eq:SE if}) in discussions of the pseudogap in the single-particle spectral function if impurity scattering is reasonably weak. 

Panel (b) of Figure \ref{fig:Acompare} shows $A_{if}(\mathbf{p},\omega)$ for several $\mathbf{p}$ (see Fig. \ref{fig:fsplots}), including a hot spot (solid trace, black online), a point $\mathbf{p}_c$ which would be at the center of the ``hole pocket" in the SDW state (dashed trace, green online), a momentum $\mathbf{p}_1$ far from the hot spot but on the noninteracting Fermi surface and a momentum $\mathbf{p}_2$ which would be near the back side of the hole pocket. At the hot spot, one observes two peaks, symmetrically disposed around the chemical potential. At the center of the ``hole-pocket", one also sees two identical peaks, but this time not centered at the chemical potential. At the other two momenta, one sees a large peak indicative of a conventional Fermi liquid quasiparticle and a small `shadow peak' at the location of the other quasiparticle state. All of these features may be understood in terms of a broadening of the mean-field solution.

\begin{figure}
\centering
\includegraphics[width=0.9\textwidth]{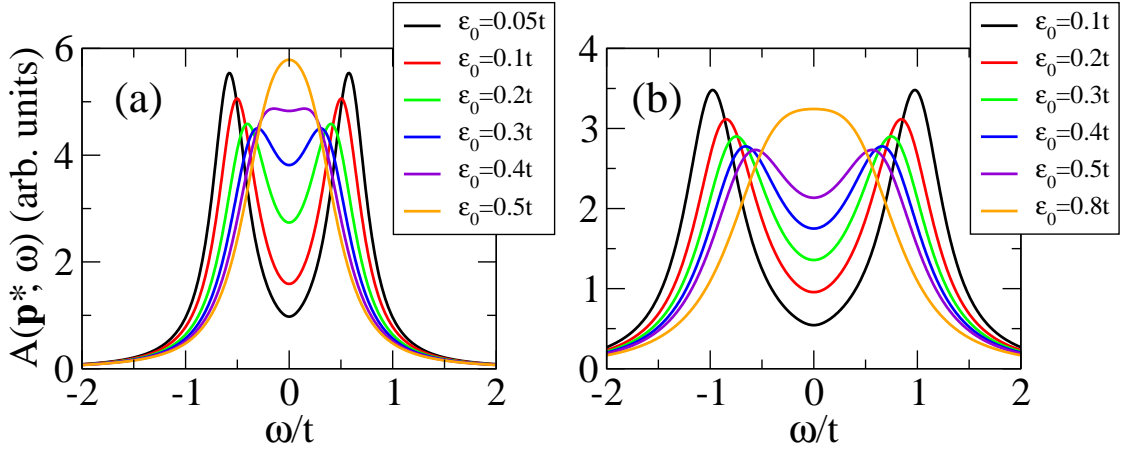}
\caption{\small{The spectral function $A(\mathbf{p}^\ast,\omega)$ at the hot spot $\mathbf{p}^\ast$ for $1/\tau=0.1t$ and $\mu=0.175t$. (a): $\lambda/t^2=0.1$ and different values of $\varepsilon_0$. (b): $\lambda/t^2=0.3$ and different values of $\varepsilon_0$.} 
}
\label{fig:Aplots}
\end{figure}

In Figure \ref{fig:Aplots} we investigate the dependence of the spectral function on the parameters $\lambda$ and $\varepsilon_0$. Each panel shows the spectral function at the hot spot, computed for a fixed $\lambda$ and several different $\varepsilon_0$. The gap scale (defined from the peak to peak distance at the smallest $\varepsilon_0$) increases with increasing $\lambda$. As $\varepsilon_0$ is increased, the low-energy density of states increases (gap fills in) and at larger $\varepsilon_0$, the gap magnitude (defined from the peak separation, when visible) decreases, but at a rate slower than the increase of the low-energy density of states. Thus the suppression of the pseudogap has more to do with the gap filling in than with a gap decrease. 

Comparing our results to data suggests that it is reasonable to associate an increase in temperature with an increase in the parameter $\varepsilon_0$ (\textit{i.e.} a decrease in the correlation length $\xi$ as observed in Ref [\onlinecite{Motoyama07}]), while the increase of $\lambda$ produces effects similar to those observed in electron-doped cuprates when doping is decreased. To qualitatively relate theory to experiment, we therefore fix the chemical potential and model changes in doping by changes in $\lambda$ and changes in temperature by changes in $\varepsilon_0$.


\section{Current vertex function and frequency-dependent longitudinal  conductivity}
\label{conductivity}

The longitudinal conductivity $\sigma_{xx}$ is given in terms of the polarization function $\Pi$ as
\begin{equation}
\sigma_{xx}(i\Omega_n)=\frac{\Pi_P(i\Omega_n)+\Pi_D}{\Omega_n}
\label{sigdef1}
\end{equation}
where the paramagnetic $P$ and diamagnetic $D$ contributions to the polarization function are given in terms of the current vertex $\Gamma^J$ as 
\begin{equation}
\Pi_P(i\Omega_n)=2\sigma_Q \lim_{q\rightarrow 0}T\sum_{\epsilon_n}\int \frac{d\mathbf{p}}{(2\pi)^2} v_p^x G(\mathbf{p},i\epsilon_n)\Gamma^J_x(\mathbf{p},\mathbf{p}+\mathbf{q},i\epsilon_n,i\epsilon_n+i\Omega_n)G(\mathbf{p}+\mathbf{q},i\epsilon_n+i\Omega_n),
\label{Pip}
\end{equation}
where $\sigma_Q=e^2/\hbar$ is the conductance quantum and $v_p^x=\partial \varepsilon_p/\partial p_x$, and 
\begin{equation}
\Pi_D=2\sigma_QT\sum_{\epsilon_n}\int \frac{d\mathbf{p}}{(2\pi)^2}\varepsilon_p^{xx}G(\mathbf{p},i\epsilon_n),
\label{Pid}
\end{equation}
where $\varepsilon_p^{xx}=\partial^2 \varepsilon_p/\partial p_x^2$. 

The magnitude of the current vertex function ${\vec\Gamma}^J$ is related to the relative sizes of the frequency and momentum dependence of the self energy. We have seen in the previous section that the momentum dependence of the self energy is not negligible and thus expect the current vertex correction to be important. An important constraint on calculations is the Ward identity following from current conservation; this ensures that  the conductivity obeys the ``$f$-sum'' rule. The Ward identity relates the density vertex $\Gamma^\rho$ and current vertices ${\vec \Gamma}^J$ to the electron propagator via 
\begin{equation}
G^{-1}(\mathbf{p}+\mathbf{q},i\epsilon+i\Omega)-G^{-1}(\mathbf{p},i\epsilon)=i\Omega\Gamma^\rho(\mathbf{p},\mathbf{p}+\mathbf{q},i\epsilon,i\epsilon+i\Omega)-\mathbf{q}\cdot {\vec \Gamma}^J(\mathbf{p},\mathbf{p}+\mathbf{q},i\epsilon,i\epsilon +i\Omega).
\label{GWI}
\end{equation}
Taking the $\mathbf{q} \to 0$ limit with $\Omega$ fixed to $0$ gives 
\begin{equation}
\lim_{q\to 0}{\vec \Gamma}^J(\mathbf{p},\mathbf{p}+\mathbf{q},i\epsilon,i\epsilon)=\mathbf{v}_p+\frac{\partial \Sigma(\mathbf{p},i\epsilon)}{\partial \mathbf{p}}.
\label{GWI-J}
\end{equation}

To obtain the vertex function, we follow the procedure  outlined in Refs [\onlinecite{Schrieffer,Koba51}]: insert the free vertex on each bare electron line in the diagrammatic expansion of the electron Green's function, and then amputate the resulting diagrams. The diagrammatic expansion for the current vertex function is shown in Fig \ref{fig:PT diagrams} (b), and the corresponding analytic expression is  (henceforth we drop the superscript $J$ and remove one of the two momentum arguments because we deal only with the current vertex in the ${\mathbf q}\rightarrow 0$ limit) 
\begin{multline}\label{eq:VC eq}
\Gamma_x(\mathbf{p},i\epsilon_n,i\epsilon_n+i\Omega_n)
=v_p^x+g^2T\int\frac{d\mathbf{q}}{(2\pi)^2}\frac{\chi_0}{\xi^{-2}+(\mathbf{q-\mathbf{Q}})^2}\\
\times
G_0(\mathbf{p}+\mathbf{q},i\epsilon_n)G_0(\mathbf{p}+\mathbf{q},i\epsilon_n+i\Omega_n)v_{p+q}^x.
\end{multline}
Approximating $v_{p+q}\approx v_{p+Q}$, Eq (\ref{eq:VC eq}) is evaluated as 
\begin{equation}\label{eq:VC}
\Gamma_x(\mathbf{p},i\epsilon_n,i\epsilon_n+i\Omega_n)=v_p^x+\frac{\Sigma(\mathbf{p},i\epsilon_n)-\Sigma(\mathbf{p},i\epsilon_n+i\Omega_n)}{i\Omega_n+i[\mathrm{sgn}(\epsilon_n+\Omega_n)-\mathrm{sgn}\epsilon_n]/2\tau}v_{p+Q}^x,
\end{equation}
which is consistent with Eq (\ref{GWI-J}) because Eq (\ref{eq:SE}) shows that $\partial \Sigma(\mathbf{p},i\epsilon_n)/\partial p_\alpha=-v_{p+Q}^\alpha\partial \Sigma(\mathbf{p},i\epsilon_n)/\partial (i\epsilon_n)$. We therefore conclude that Eq (\ref{eq:VC}) for the current vertex function is a conserving approximation. 

In the study of the fluctuation conductivity near the superconducting transition, the Aslamazov-Larkin (AL) contribution is usually important (see \textit{e.g.} Ref [\onlinecite{Larkin}]). Its calculation requires the inclusion of the dynamic term in the spin-fluctuation propagator $\chi(\mathbf{q},\omega)$. We calculated the AL contribution using Eq (\ref{eq:chi}), and found that it is negligible, because $\chi(\mathbf{q},\omega)$ is peaked at a finite momentum $\mathbf{Q}$. 

The physics of the vertex correction may be understood by comparison to the mean-field solution in the ordered state. To demonstrate the main issues with a minimum of notational complexity, we discuss the one-dimensional model, in which we linearize the dispersion about the Fermi energy, measure momenta from the Fermi momentum, and assume the ordering wave vector $Q=2k_F$. The mean-field solution is characterized by normal ($G_{MF}\sim <c^\dagger_pc_p>$) and anomalous ($F\sim <c^\dagger_pc_{p+Q}>$) Green's functions given for right ($a=+$) and left ($a=-$) moving electrons by 
\begin{eqnarray}
G_{MF}^a(p,i\omega)&=&-\frac{i\omega+avp}{\omega^2+v^2p^2+\Delta^2},
\\
F_{MF}^a(p,i\omega)&=&-\frac{\Delta}{\omega^2+v^2p^2+\Delta^2},
\end{eqnarray}
and the conductivity is given schematically by 
\begin{equation}
\sigma\propto\frac{1}{\Omega}Tr\left[GG-FF\right]
\label{sigmaMF},
\end{equation}
with the trace over frequency, momentum and left/right index $a$.

Turning now to the theory in the fluctuation regime, we have (for $\omega>0$) 
\begin{equation}
G_{LRA}(p,i\omega)=-\frac{i\omega+vp+\frac{i}{2\tau}+i\varepsilon_0}{\omega^2+v^2p^2+\Delta^2-i[(i\omega-vp)(\frac{1}{2\tau}+\varepsilon_0)+\frac{1}{2\tau}(i\omega+vp)]}.
\label{GLRA2}
\end{equation}
The absence of long-range order means that $F=0$, so that $\sigma$ is evaluated directly from Eq (\ref{sigdef1}) while use of Eq (\ref{eq:VC}) for the vertex function and Eq (\ref{eq:SE 1d}) gives 
\begin{equation}
\Gamma^{++}=v\left(1-\frac{\Delta^2}{\left(i\omega+i\Omega+vp+i\left(\frac{1}{2\tau}+\varepsilon_0\right)\right)\left(i\omega+vp+i\left(\frac{1}{2\tau}+\varepsilon_0\right)\right)}
\right),
\end{equation}
if $\mathrm{sgn}( \omega+\Omega)=\mathrm{sgn}(\omega)=+$, and 
\begin{equation}
\Gamma^{-+}=v\left(1-\frac{\Delta^2\left(1+\frac{2\varepsilon_0}{\Omega+\frac{1}{\tau}}\right)}{\left(i\omega+i\Omega+vp+i\left(\frac{1}{2\tau}+\varepsilon_0\right)\right)\left(i\omega+vp-i\left(\frac{1}{2\tau}+\varepsilon_0\right)\right)}\right),
\end{equation}
if $\mathrm{sgn}( \omega+\Omega)=+$ but $\mathrm{sgn}(\omega)=-$. 

Substituting into Eq (\ref{sigdef1}), we see that the first of the two terms in the vertex function reproduces the $GG$ term. The second of the two terms reproduces the $FF$ contribution, which, in the ordered state, carries the coherence factors which for example distinguish antiferromagnetism from superconductivity. Thus the vertex correction does what is  required to produce the correct form of the near gap conductivity. However, we see that in addition, in the physically crucial $\mathrm{sgn}(\omega+\Omega)\neq \mathrm{sgn}(\omega)$ regime there is an extra term, of order $\varepsilon_0/(\Omega+1/\tau)$ which diverges as $\Omega+1/\tau \to 0$ but is unimportant for $|\Omega+1/\tau|>\varepsilon_0$. The structure of this term is a defect of the Lee-Rice-Anderson approximation. We believe it occurs because this theory produces an incorrect form for the subgap density of states, which should vanish as frequency $\omega\rightarrow 0$. Indeed in the one-dimensional case it is known that the low-frequency density of states is due to amplitude singularities in the flucuating order parameter, which become exponentially rare at low frequencies.\cite{Monien01}

The problem can also be cured by a self-consistent treatment such as FLEX, but this is known to give an incorrect form for the pseudogap density of states.\cite{Monien01}  We have not been able to identify a consistent and physically reasonable cure for the divergence which is applicable also in two dimensions, so we adopt the expedient of introducing an impurity scattering which cuts off the divergence. We shall see, however, that the theory can still produce an unphysical dip in the low-frequency conductivity. 

Returning to the two-dimensional model of primary interest in this paper, we combine Eqs. (\ref{sigdef1},\ref{Pip},\ref{Pid},\ref{eq:VC}), perform the analytical continuation, and obtain 
\begin{equation}\label{eq:conductivity}\mathrm{Re}\sigma_{xx}(\Omega)=\mathrm{Re}\sigma_{xx}^{(I)}(\Omega)+\mathrm{Re}\sigma_{xx}^{(II)}(\Omega)+\mathrm{Re}\sigma_{xx}^{(III)}(\Omega),\end{equation}
where 
\begin{equation}
\label{eq:conductivity 1}
\mathrm{Re}\sigma_{xx}^{(I)}(\Omega)=\sigma_Q\int\frac{d\mathbf{p}}{(2\pi)^2}\bigl((v_p^x)^2-v_p^xv^x_{p+Q}\bigr)\int_{-\infty}^\infty \frac{d\omega}{2\pi}\frac{f(\omega)-f(\omega+\Omega)}{\Omega}A(\mathbf{p},\omega)A(\mathbf{p},\omega+\Omega),
\end{equation}
\begin{multline}\label{eq:conductivity 2}
\mathrm{Re}\sigma_{xx}^{(II)}(\Omega)=\sigma_Q\frac{1/\tau}{\Omega^2+1/\tau^2}\int \frac{d\mathbf{p}}{(2\pi)^2}v^x_pv^x_{p+Q}
\\
\times\int_{-\infty}^\infty\frac{d\omega}{2\pi}\frac{f(\omega)-f(\omega+\Omega)}{\Omega}\bigl[A(\mathbf{p},\omega)+A(\mathbf{p},\omega+\Omega)\bigr],
\end{multline}
and 
\begin{multline}\label{eq:conductivity 3}
\mathrm{Re}\sigma_{xx}^{(III)}(\Omega)=2\sigma_Q\frac{1/\tau^2}{\Omega^2+1/\tau^2}\int \frac{d\mathbf{p}}{(2\pi)^2}v^x_pv^x_{p+Q}\\
\times\int_{-\infty}^\infty\frac{d\omega}{2\pi}\frac{f(\omega)-f(\omega+\Omega)}{\Omega}\frac{\mathrm{Re}G(\mathbf{p},\omega+\Omega)-\mathrm{Re}G(\mathbf{p},\omega)}{\Omega},
\end{multline}
where $f(x)$ is the Fermi function. In the calculation, we assume that the most important effect of temperature is on $\xi$ (or $\varepsilon_0$), and neglect thermal broadening of the Fermi function. As a result, $f(x\leq 0)=1$ and $f(x>0)=0$. In the limit $\tau\to \infty$, $\mathrm{Re}\sigma_{xx}^{(I)}$ remains finite, $\mathrm{Re}\sigma_{xx}^{(II)}=\Delta S\delta(\Omega)$, and $\mathrm{Re}\sigma_{xx}^{(III)}(\Omega)\to 0$. Thus, $\mathrm{Re}\sigma_{xx}^{(II)}$ is the divergence discussed above. Its weight $\Delta S$ is given by 
\begin{equation}\label{eq:deltaS}
\Delta S=2\pi \sigma_Q\int \frac{d\mathbf{p}}{(2\pi)^2}\int \frac{d\omega}{2\pi}\Bigl(-\frac{d f(\omega)}{d\omega}\Bigr)v_p^x v_{p+Q}^xA(\mathbf{p},\omega).
\end{equation}
We find that $\Delta S<0$ for the band dispersion appropriate to cuprates, since $v_p^x v_{p+Q}^x<0$ in most part of the Brillouin zone where $A(\mathbf{p},0)$ is appreciable. 

\begin{figure}
\centering
\includegraphics[width=.9\textwidth]{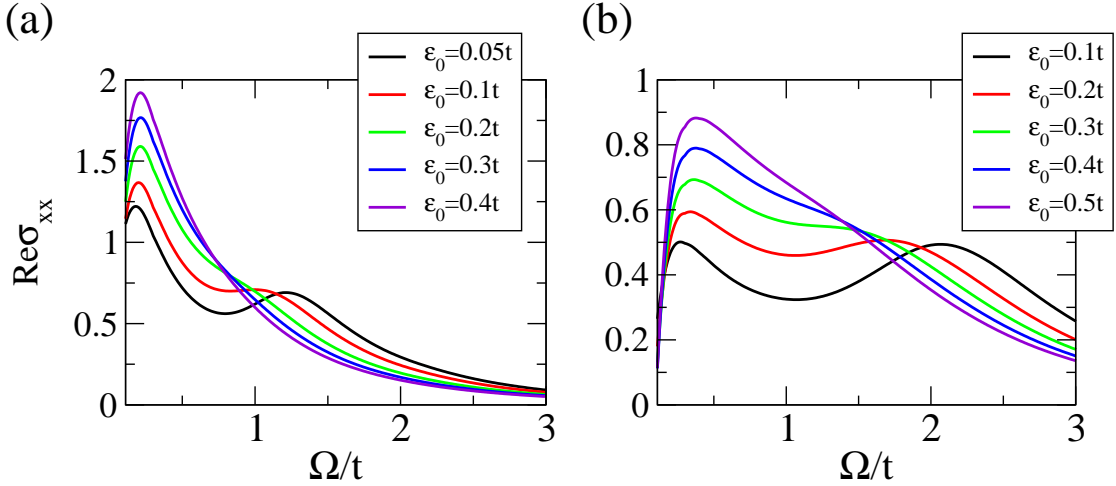}
\caption{\small{The longitudinal conductivity per plane evaluated from Eq (\ref{eq:conductivity}) with $1/\tau=0.1t$ and $\mu=0.175t$. (a): $\lambda/t^2=0.1$ and different $\varepsilon_0$. (b): $\lambda/t^2=0.3$ and different $\varepsilon_0$. To convert to physical units, one must multiply the calculated result by the conductance quantum, $\sigma_Q=e^2/\hbar$, and divide it by the inter-plane distance.}}
\label{fig:conductivity}
\end{figure}

Figure \ref{fig:conductivity} shows $\mathrm{Re}\sigma_{xx}(\Omega)$ calculated from Eq (\ref{eq:conductivity}) with $1/\tau=0.1t$ and $\mu=0.175t$ for different values of $\lambda$ and $\varepsilon_0$. All curves in Figure \ref{fig:conductivity} show anomalous low-frequency behavior, arising from Eq (\ref{eq:conductivity 2}). Since $\Delta S<0$, the DC limit can be made negative (not shown here) for larger values of $\lambda$ or $\tau$. As $\Omega$ increases, this anomalous contribution is quickly suppressed due to the prefactor $\frac{1/\tau}{\Omega^2+1/\tau^2}$. In Figure \ref{fig:conductivity}, we see that $\mathrm{Re}\sigma_{xx}(\Omega)$ behaves as expected for $\Omega\gtrsim 3/\tau$, and we shall concentrate on this regime. In this regime, for small $\varepsilon_0$ ($\varepsilon_0/t=0.05,0.1,0.2$ in panel (a) and $\varepsilon_0/t=0.1,0.2,0.3,0.4$ in panel (b)), there are peaks around $2\Delta_{pg}$ as determined from Figure \ref{fig:Aplots}. This peak structure is reminiscent of that in mean-field calculations, as shown in Figure \ref{fig:MF} (a). The peak becomes weaker for larger $\varepsilon_0$ (smaller $\xi$). We now use the association of $\lambda$ and $\varepsilon_0$ with $x$ and $T$ as discussed in Sec. \ref{model} to relate these results to experimental observations. Comparing the two panels in Figure \ref{fig:conductivity} suggests that at low temperatures, $\mathrm{Re}\sigma_{xx}$ has an optical peak, the peak position decreases with doping, the peak vanishes at some temperature $T^\ast$, $T^\ast$ increases with underdoping, and for fixed doping, there is a spectral weight transfer from high-frequency region to low-frequency region, as $T$ is increased.

We have verified numerically that the calculated conductivity obeys the $f$-sum rule
\begin{equation}\label{eq:sumrule}
\int_0^\infty d\Omega\mathrm{Re}\sigma_{xx}(\Omega)=\frac{\pi}{2}\Pi_D.
\end{equation}
To see this analytically, we consider the case of small $1/\tau$, such that $\mathrm{Re}\sigma_{xx}^{(III)}(\Omega)$ can be neglected and $\int_{-\infty}^\infty\frac{d\Omega}{\pi}\mathrm{Re}\sigma_{xx}^{(II)}(\Omega)$ can be approximated as 
\begin{equation}
\int_{-\infty}^\infty\frac{d\Omega}{\pi}\mathrm{Re}\sigma^{(II)}_{xx}(\Omega)
\approx \Delta S/\pi. 
\end{equation}
At the same time, $\int_{-\infty}^\infty\frac{d\Omega}{\pi}\mathrm{Re}\sigma^{(I)}_{xx}(\Omega)$ can be evaluated using the Kramers-Kr\"onig relation between $A(\mathbf{p},\omega)$ and $\mathrm{Re}G(\mathbf{p},\omega)$, and the result is 
\begin{equation}
\int_{-\infty}^\infty\frac{d\Omega}{\pi}\mathrm{Re}\sigma_{xx}^{(I)}(\Omega)=
-2\sigma_Q\int\frac{d\mathbf{p}}{(2\pi)^2}v_p^x(v_p^x-v_{p+Q}^x)\int_{-\infty}^\infty\frac{d\epsilon}{\pi}
f(\epsilon)A(\mathbf{p},\epsilon)\mathrm{Re}G(\mathbf{p},\epsilon).
\end{equation}
$\Pi_D$ is found to be equal to the sum of the above two equations. Since $\mathrm{Re}\sigma_{xx}(\Omega)$ is an even function of $\Omega$, we obtain Eq (\ref{eq:sumrule}).

The conductivity $\sigma_{xx}(\Omega)$ in the mean-field theory of the SDW state (Eq (\ref{eq:conductivity MF})) can be obtained in the present leading-order perturbation theory, by simply substituting Eq (\ref{eq:SE KS}) into various Green's functions in Eq (\ref{eq:conductivity}). The vertex corrections are crucial in this derivation; if we had neglected the vertex corrections, we would effectively have neglected the off-diagonal terms in Eq (\ref{eq:G MF}) (the terms proportional to $\Delta$). The conductivity in the mean-field SDW state for $\Delta=0.3t$ (solid line) and $\Delta=0.6t$ (dashed line) is shown in Fig \ref{fig:MF} (a). The low-frequency parts of these curves are well described by a Drude peak without any anomalous dip. One possible reason is that in the limit $\tau\to\infty$, both $A$ and $G^R$ in the mean-field theory are singular, unlike Eqs (\ref{eq:SE}, \ref{eq:G}) which are finite due to scattering from spin fluctuations. 


\section{Hall conductivity in the perturbation theory}
\label{Hall}

\begin{figure}[htbp]
\centering
\includegraphics[width=0.8\textwidth]{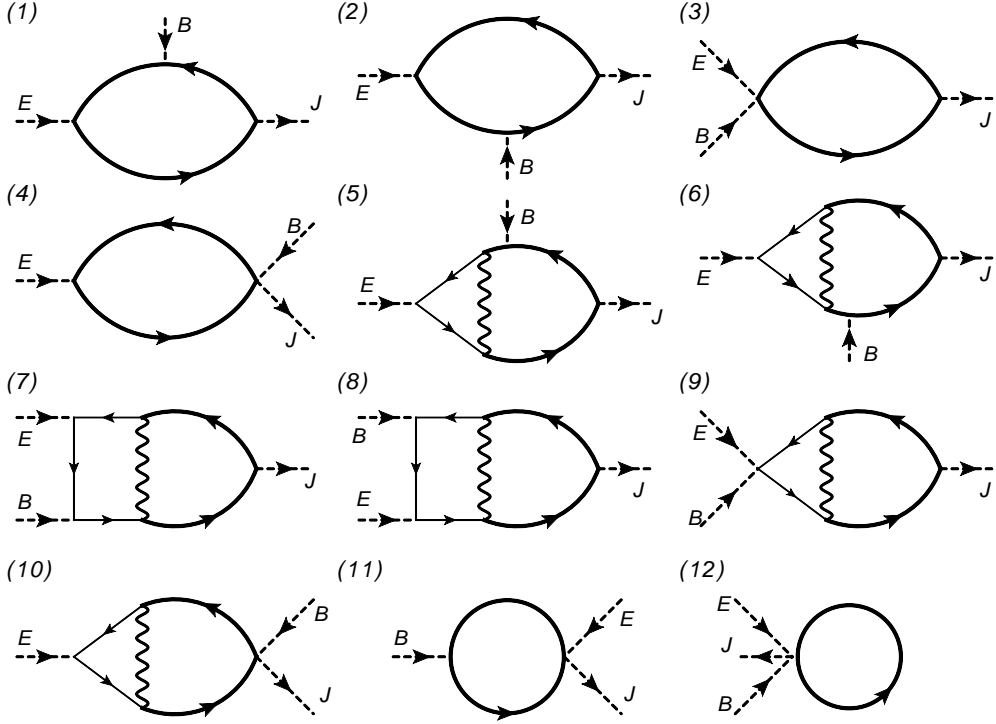}
\caption{\small{Diagrams for the Hall conductivity. Thick solid lines represent the dressed Green's function $G(\mathbf{p},i\epsilon_n)$, thin solid lines represent the bare Green's function $G_0(\mathbf{p},i\epsilon_n)$, and wavy lines represent the spin-fluctuation propagator. The vertices with one dashed line are associated with $v_p^\alpha=\partial\varepsilon_p/\partial p_\alpha$, the vertices with two dashed lines are $\varepsilon_p^{\alpha\beta}=\partial^2\varepsilon_p/\partial p_\alpha \partial p_\beta$, and the vertices with three dashed lines are $\varepsilon_{p}^{\alpha\beta\gamma}=\partial^3\varepsilon_p/\partial p_\alpha \partial p_\beta\partial p_\gamma$. The Greek indices, $\alpha, \cdots$, refer to directions of the external fields, $\mathbf{E}$, $\mathbf{J}$, and $\mathbf{B}$. }}
\label{fig:Hall PT diagrams}
\end{figure}

In this section, we develop the formalism for calculating the Hall conductivity $\sigma_{xy}$ in the leading order perturbation theory. The calculation of $\sigma_{xy}$ in the self-consistent Born approximation can be found in Refs [\onlinecite{Fukuyama69,Fukuyama69-2,Kohno88}]. At the level of approximation employed here, we find it easier to apply the method developed in Ref [\onlinecite{Altshuler}] to the conductivity diagrams shown in Fig \ref{fig:PT diagrams} (c). Rewriting the diagrams in Fig \ref{fig:PT diagrams} (c) in terms of the bare Green's functions, replacing every electron momentum $\mathbf{p}$ in the loop, according to the minimal coupling rule, by $\mathbf{p}-\frac{e}{c}\mathbf{A}$, and expanding the resulting diagrams to first order in $\mathbf{A}$, we obtain the diagrams shown in Figure \ref{fig:Hall PT diagrams}, in which the intersections where the magnetic field lines denoted by $B$ meet the dressed Green's functions $G(\mathbf{p},i\epsilon_n)$ represented by thick solid lines are the dressed magnetic vertices, which are calculated according to Fig \ref{fig:PT diagrams} (b).

Summing all the diagrams in Figure \ref{fig:Hall PT diagrams} and expanding the resulting expression up to first order in $\mathbf{k}$, we find that (1) the terms independent of $\mathbf{k}$ vanish, and (2) the terms of first order in $\mathbf{k}$ depends on $\mathbf{B}=i\mathbf{k}\times \mathbf{A}=B\hat{\mathbf{z}}$, signaling gauge invariance. The Hall conductivity on the Matsubara axis can be expressed as 
\begin{equation}\label{eq:sigmaxy}
\sigma_{xy}(i\Omega_n)=\frac{\pi}{2}\sigma_Q\frac{Ba^2}{\Phi_0}\frac{1}{i\Omega_n}\int \frac{d\mathbf{p}}{(2\pi)^2} \Bigl\{S_1-S_2+S_3+S_4-S_5\Bigr\},
\end{equation}
where $\Phi_0=\pi\hbar c/e$ is the superconducting flux quantum, 
\begin{equation}
S_1=T\sum_{i\epsilon_n}\bigl(\partial_y \Gamma_p^x v_p^y -\partial_y \Gamma_p^yv_p^x\bigr)\bigl(\mathcal{G}\partial_x\mathcal{G}(+)-\mathcal{G}(+)\partial_x\mathcal{G}\bigr),
\end{equation}
\begin{equation}
S_2=T\sum_{i\epsilon_n} \bigl(\partial_x\Gamma_p^x v_p^y-\partial_x\Gamma_p^y v_p^x\bigr)\bigl(\mathcal{G}\partial_y\mathcal{G}(+)-\mathcal{G}(+)\partial_y\mathcal{G}\bigr),
\end{equation}
\begin{equation}
S_3=T\sum_{i\epsilon_n} \bigl(\Gamma_p^x v_p^y -\Gamma_p^y v_p^x\bigr)\bigl(\partial_x\mathcal{G}(+)\partial_y\mathcal{G}-\partial_x\mathcal{G}\partial_y\mathcal{G}(+)\bigr),
\end{equation}
\begin{equation}
S_4=g^2T^2\sum_{i\epsilon_n}v_p^y\mathcal{G}\mathcal{G}(+)\int \frac{d\mathbf{q}}{(2\pi)^2}\chi(\mathbf{p}-\mathbf{q})(\varepsilon_q^{xx}v_q^y-\varepsilon_q^{xy}v_q^x)
\bigl(\mathcal{G}_0(+)\mathcal{G}_0^2-\mathcal{G}_0\mathcal{G}_0(+)^2\bigr),
\end{equation}
and 
\begin{equation}
S_5=g^2T^2\sum_{i\epsilon_n}v_p^x\mathcal{G}\mathcal{G}(+)\int \frac{d\mathbf{q}}{(2\pi)^2}\chi(\mathbf{p}-\mathbf{q})(\varepsilon_q^{xy}v_q^y-\varepsilon_q^{yy}v_q^x)
\bigl(\mathcal{G}_0(+)\mathcal{G}_0^2-\mathcal{G}_0\mathcal{G}_0(+)^2\bigr).
\end{equation}
In writing these equations, we have used short-hand notations: $\mathcal{G}=G(\mathbf{p},i\epsilon_n)$, $\mathcal{G}(+)=G(\mathbf{p},i\epsilon_n+i\Omega_n)$, $\Gamma_p^x=\Gamma_x(\mathbf{p},i\epsilon_n,i\epsilon_n+i\Omega_n)$, $\mathcal{G}_0=G_0(\mathbf{q},i\epsilon_n)$, and $\mathcal{G}_0(+)=G_0(\mathbf{q},i\epsilon_n+i\Omega_n)$. Using the spin-fluctuation propagator in the Kampf-Schrieffer model, Eq (\ref{eq:chi KS}), after a lengthy calculation, we can show that Eq (\ref{eq:sigmaxy}) reduces to that in the mean-field theory of the SDW state (Appendix \ref{MF}).\cite{Zimmers07} The proper treatment of the vertex functions as discussed here is crucial in arriving at this conclusion.

The frequency summation in Eq (\ref{eq:sigmaxy}) is standard.\cite{Mahan} The physical observable $\sigma_{xy}(\Omega)$ is obtained by analytical continuation $i\Omega_n\to\Omega+i\delta$. The expression for $\mathrm{Im}\sigma_{xy}(\Omega)$ is quite cumbersome. Here, we focus on the results, leaving detailed expressions to Appendix \ref{sxy}. 

\begin{figure}
\centering
\includegraphics[width=.9\textwidth]{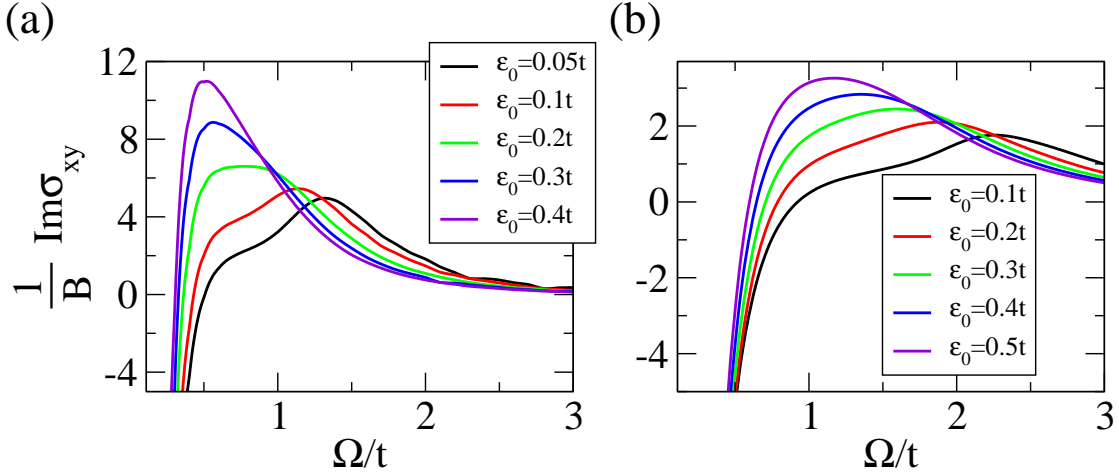}
\caption{\small{The Hall conductivity per plane divided by $B$, $\mathrm{Im}\sigma_{xy}(\Omega)/B$, calculated from Eq (\ref{eq:Hall}) with $1/\tau=0.1t$ and $\mu=0.175t$. (a): $\lambda/t^2=0.1$ and different values of $\varepsilon_0$. (b): $\lambda/t^2=0.3$ and different values of $\varepsilon_0$. To convert to physical units, one must multiply the calculated result by the conductance quantum and the in-plane unit cell area, and divide it by the superconducting flux quantum, $\Phi_0=hc/2e$, and the inter-plane distance.}}
\label{fig:Hall}
\end{figure}

\begin{figure}
\centering
\includegraphics[width=.9\textwidth]{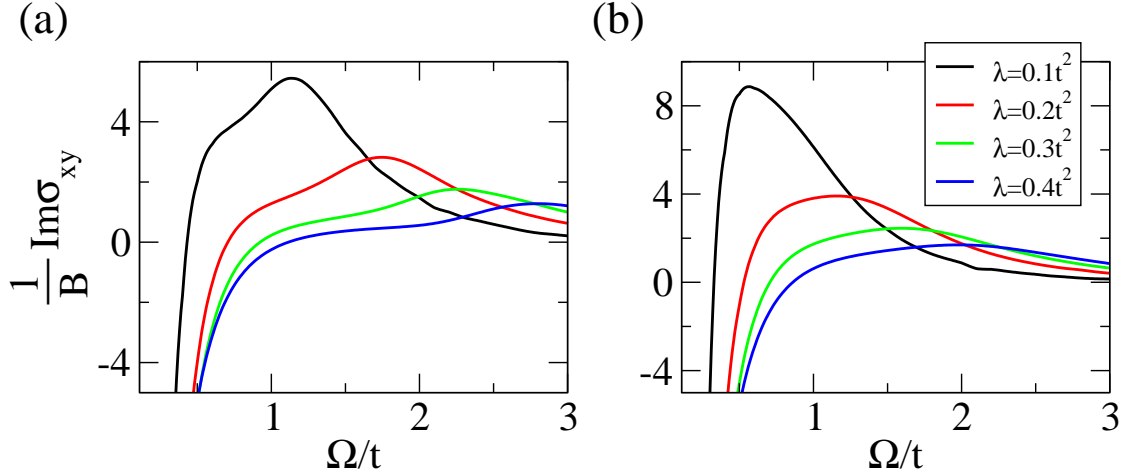}
\caption{\small{The Hall conductivity per plane divided by $B$, $\mathrm{Im}\sigma_{xy}(\Omega)/B$, calculated from Eq (\ref{eq:Hall}) with $1/\tau=0.1t$ and $\mu=0.175t$. (a): $\varepsilon_0/t=0.1$. (b): $\varepsilon_0/t=0.3$. To convert to physical units, see Figure \ref{fig:Hall}.}}
\label{fig:Hall e0}
\end{figure}

Figures \ref{fig:Hall} and \ref{fig:Hall e0} show $\mathrm{Im}\sigma_{xy}(\Omega)$ calculated from Eq (\ref{eq:Hall}) for $1/\tau=0.1t$, $\mu=0.175t$, and various values of $\lambda$ and $\varepsilon_0$. The frequency scales in Figures \ref{fig:Hall} and \ref{fig:Hall e0} are selected to highlight the frequency window most relevant to experiments\cite{Zimmers07} where $\sigma_{xy}$ was measured for $\Omega>0.3t\approx 1000\mathrm{cm}^{-1}$. The low-frequency part  ( $\Omega \lesssim 1/\tau$) of $\mathrm{Im}\sigma_{xy}(\Omega)$ suffers from the same difficulty as does $\mathrm{Re}\sigma_{xx}(\Omega)$ due to the perturbative nature of the calculation and we do not show results in this region. 

Comparing to Ref [\onlinecite{Zimmers07}] suggests that the present calculation captures important features of data. First, $\mathrm{Im}\sigma_{xy}$ can be made negative at low frequencies, although we start with a single-band model with a hole-like Fermi surface. A previous study suggested that the appearance of the negative $\mathrm{Im}\sigma_{xy}$ in electron-doped cuprates is a signature of the long-range spin-density wave order.\cite{Zimmers07} Our findings here suggest that fluctuating order can also explain this behavior. The vertex corrections shown in Figure \ref{fig:Hall PT diagrams} are important for this conclusion.   We found that $\mathrm{Im}\sigma_{xy}$ remains positive in the entire frequency range, if only the diagrams 1-4, 11, and 12 in Figure \ref{fig:Hall PT diagrams} are kept and the vertex corrections to magnetic vertices are neglected. Kontani and co-workers, using the FLEX approximation, also emphasized the importance of the magnetic field vertex corrections.\cite{Kontani99} As shown in Figure \ref{fig:Hall}, $\mathrm{Im}\sigma_{xy}(\omega)$ has relatively sharp peaks for small $\varepsilon_0$ ($\varepsilon_0/t=0.05,0.1$ in panel (a), and $\varepsilon_0/t=0.1,0.2$ in panel (b)) around $2\Delta_{pg}$, showing precursor effect to that obtained from a mean-field calculation shown in Figure \ref{fig:MF} (b). The peak gradually vanishes as  $\varepsilon_0$ increases. Furthermore, fixing $\lambda$ (or doping $x$) and increasing $\varepsilon_0$ (or temperature $T$), $\mathrm{Im}\sigma_{xy}(\Omega)$ increases from negative to positive  at low frequencies and decreases at high frequencies. This is qualitatively consistent with the trend observed experimentally in electron-doped cuprates in the underdoped regime.\cite{Zimmers07}   From Figure \ref{fig:Hall e0}, we see that at fixed $\varepsilon_0$, $\mathrm{Im}\sigma_{xy}(\Omega)$ decreases with increasing $\lambda$ (or decreasing $x$) from positive to negative at low frequencies, and increases with increasing $\lambda$ at high frequencies, again qualitatively consistent with data.\cite{Zimmers07} However, in our study, $\mathrm{Im}\sigma_{xy}(\Omega)$ remains positive at high frequencies, inconsistent with data.\cite{Zimmers07} More quantitatively, if (following the discussion of the longitudinal conductivity above) we assume that $\lambda/t^2=\varepsilon_0/t=0.1$ is a reasonable representation of cuprates at $0.12$ electron doping, we see that the predicted zero crossing in $\sigma_{xy}$ occurs at $\Omega\sim 0.15-0.2$eV, again semiquantitatively consistent with data. However, our calculation exhibits more temperature dependence than is found in data.


\section{summary}
\label{summary}

In this paper, we used the Lee-Rice-Anderson model to study two-dimensional electrons scattered from static antiferromagnetic spin fluctuations, with potential applications to electron-doped cuprates in the underdoped regime where the long-range spin-density wave ground state is expected. Our theory is in a sense complementary to that of Kontani \textit{et al} who used a fluctuation-exchange approximation most applicable in the overdoped region.\cite{Kontani99,Jenkins10} The theory has two important parameters: $\lambda$ which controls the gap amplitude, and $\varepsilon_0$ which represents the effect of non-zero temperature. 

We first discussed single-particle properties with the self-energy calculated in the leading-order perturbation theory. There is pseudogap opening for relatively small $\varepsilon_0$ which is related to the spin correlation length $\xi$ via $\varepsilon_0=v_F/\xi$ where $v_F$ is the Fermi velocity. As $\varepsilon_0$ increases, the pseudogap is gradually filled in with a moderate change in the size of the pseudogap. The value of the pseudogap is primarily determined by the coupling constant $\lambda$ between electrons and spin fluctuations. We assume that $\lambda$ is increasing as underdoping, and $\varepsilon_0$ is increasing as increasing the temperature. The conductivity $\mathrm{Re}\sigma_{xx}(\Omega)$ is calculated in a conserving approximation which respects the $f$-sum rule. The current vertex has unphysical low-energy features. We found that in order to obtain a finite DC conductivity, it is necessary to include impurity scattering. However, even with impurities, the low-frequency part of $\mathrm{Re}\sigma_{xx}(\Omega)$ still behaves anomalously. As frequency $\Omega$ increases larger than $1/\tau$, the anomalous contribution is quickly suppressed. $\mathrm{Re}\sigma_{xx}(\Omega)$ is characterized by a peak around twice the pseudogap value for relatively small $\varepsilon_0$ (large $\xi$). This is reminiscent of the peak in mean-field calculations for the long-range spin-density wave ordered state. For fixed $\lambda$ (or doping), there is a spectral weight transfer from the high-frequency region to the low-frequency region as increasing $\varepsilon_0$ (or decreasing $\xi$). For the Hall conductivity $\mathrm{Im}\sigma_{xy}(\Omega)$, we focused on the experimentally accessible frequency regime $\Omega>0.3t$, and showed that $\mathrm{Im}\sigma_{xy}(\Omega)$ can be either positive or negative at small frequencies, depending on parameters $\lambda$ and $\varepsilon_0$. A negative $\mathrm{Im}\sigma_{xy}(\Omega)$ is rather non-trivial, and is a consequence of current vertex corrections.\cite{Kontani99,Jenkins10} For small $\varepsilon_0$, $\mathrm{Im}\sigma_{xy}$ has a peak structure, reminiscent of the mean-field calculations. For fixed $\lambda$ (or doping), $\mathrm{Im}\sigma_{xy}$ increases at low frequencies and decreases at high frequencies, as increasing $\varepsilon_0$ (decreasing $\xi$, or increasing temperature). For fixed $\varepsilon_0$, $\mathrm{Im}\sigma_{xy}$ increases at low frequencies and decreases at high frequencies, as decreasing $\lambda$ (or increasing doping). 

In comparison to experiment, $\sigma_{xx}$ calculated in our theory is $\sim 2-3$ times larger than data in the $0.1<\Omega<0.5$eV range (Figure \ref{fig:conductivity}). We believe that this reflects the inadequate treatment of Mott correlations. Our calculated $\sigma_{xy}$ is about 5 times larger than experiment (Figure \ref{fig:Hall} and Ref [\onlinecite{Zimmers07}]). The structure, with a negative $\sigma_{xy}$ at low frequencies and a positive value at higher frequencies is qualitatively consistent with data,\cite{Zimmers07} except that in our study, $\mathrm{Im}\sigma_{xy}$ stays positive at high frequencies, unlike data.\cite{Zimmers07} Another minor point of difference is that in the data there is little temperature dependence of the zero crossing in $\sigma_{xy}$, while in the theory the zero-crossing point shifts with $\varepsilon_0$.

One advantage of the approach in the present paper is that the results for the electron spectral function, the conductivity, and the Hall conductivity are directly related to the mean-field results if the spin propagator takes the Kampf-Schrieffer form $\chi\propto \delta(\mathbf{q}-\mathbf{Q})$. However, this approach is insufficient for the study of transport properties in the low frequency limit. 

\textit{Acknowledgment} The authors thank H. D. Drew, H. Kontani, I. Aleiner, M. R. Norman, A. V. Chubukov, A. Varlamov, and C. P\'epin for helpful discussions. The work at Columbia is supported by NSF Grant No. DMR-0705847, and the work at Argonne is supported by the US DOE, Office of Science, under contract DE-AC02-06CH11357.

\appendix


\section{Renormalization of the impurity relaxation time and the spin-fermion interaction vertex}
\label{gtau}

In our model, the fermions are scattered by both spin fluctuations, characterized by the interaction vertex $g$, and impurities, characterized by the relaxation time $\tau$. One important question is to study how one of the scattering process affects the other. This issue is addressed in this Appendix. We find that both renormalizations can be neglected in the sense to be discussed below. 

\subsection{Renormalization of the spin-fermion interaction vertex}
\label{g}

\begin{figure}
\centering
\includegraphics[width=.6\textwidth]{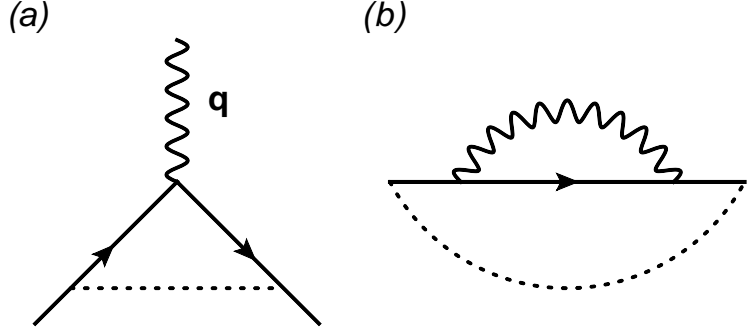}
\caption{\small{(a): the leading order diagram for the renormalization of the spin-fermion interaction vertex $g$ due to impurity scattering. (b): the leading order correction to the impurity relaxation time due to spin-fermion interaction. The dashed line is the impurity line, the solid line is the bare electron propagator $G_0$, and the wavy line is the spin-fluctuation propagator.}}
\label{fig:gtaurenorm}
\end{figure}
In this subsection, we discuss the renormalization of the spin-fermion interaction vertex $g$ in the presence of impurity scattering. The leading order correction is given by Fig \ref{fig:gtaurenorm} (a),  
\begin{equation}
\delta g/g=u^2\int \frac{d\mathbf{p}^\prime}{(2\pi)^2} G_0(\mathbf{p}^\prime,i\epsilon_n)G_0(\mathbf{p}^\prime+\mathbf{q},i\epsilon_n),
\end{equation}
where $u^2$ is the impurity potential, and we only consider the static limit for the spin-fluctuation propagator. For the momentum transfer $|\mathbf{q}-\mathbf{Q}|\sim \xi^{-1}\ll a^{-1}$, the momentum integral can be transformed to $\int d\varepsilon_p d\varepsilon_{p+Q}$, and the two integrals can be performed independently. As a result, $\delta g/g\propto u^2$, and its dependence on $i\epsilon_n$ and $\mathbf{q}$ is estimated to be $\mathcal{O}(T/\mathcal{D},a|\mathbf{q}-\mathbf{Q}|)$ which can be neglected, where $\mathcal{D}$ is the cut-off energy of the order of the fermion bandwidth. Applying the same argument to the impurity ladder diagrams, we find that the summation of the ladder diagrams gives a finite constant renormalization to $g$. As a result, we can neglect the diagrams that renormalizes $g$ by properly redefining $g$. 

\subsection{Renormalization of the impurity relaxation time $\tau$}
\label{tau}

We now discuss the renormalization of the impurity scattering relaxation time $\tau$ by the spin-fermion interaction. Figure \ref{fig:gtaurenorm} (b) shows the leading order term in calculating this renormalization, 
\begin{equation}
\delta \Bigl(\frac{1}{\tau}\Bigr)=g^2T\int \frac{d\mathbf{q}}{(2\pi)^2}\chi(\mathbf{q})\mathcal{T}(\mathbf{q}),
\end{equation}
where
\begin{equation}
\mathcal{T}(\mathbf{q})=u^2\int \frac{d\mathbf{p}^\prime}{(2\pi)^2}G_0(\mathbf{p}^\prime,i\epsilon_n)^2G_0(\mathbf{p}^\prime+\mathbf{q},i\epsilon_n).
\end{equation}
For the momentum transfer $|\mathbf{q}-\mathbf{Q}|\sim \xi^{-1}\ll a^{-1}$, the $\int d\mathbf{p}^\prime$ integral can be transformed to $\int d\varepsilon_{p^\prime}d\varepsilon_{p^\prime+Q}$. Since the integral $\int d\varepsilon_{p^\prime}$ has a double pole, this leading renormalization is negligible. This argument persists to diagrams with more spin-fluctuation lines. Thus, the renormalization of the impurity scattering relaxation time due to the spin-fermion interaction is negligible.


\section{The static approximation to Eq (\ref{eq:chi})}
\label{static}

In this Appendix, we discuss the condition under which the static spin-fluctuation propagator Eq (\ref{eq:chi static}) can be used. For simplicity, we consider the electron self-energy in the leading order perturbation theory, 
\begin{equation}
\Sigma(\mathbf{p},i\epsilon_n)=g^2T\sum_{i\omega_n}\int \frac{d\mathbf{q}}{(2\pi)^2}\chi(\mathbf{q},i\omega_n)\frac{1}{i\epsilon_n+i\omega_n-\varepsilon_{p+q}}.
\end{equation}
Substituting the spectral decomposition (see \textit{e.g.} Ref [\onlinecite{AGD}]), 
\begin{equation}
\chi(\mathbf{q},i\omega_n)=\int_{-\infty}^\infty \frac{dx}{\pi}\frac{\mathrm{Im}\chi(\mathbf{q},x)}{x-i\omega_n},
\end{equation}
where $\chi(\mathbf{q},\omega)$ is given by Eq (\ref{eq:chi}), and summing over $i\omega_n$, we obtain 
\begin{equation}\label{eq:SE W}
\Sigma(\mathbf{p},i\epsilon_n)=g^2\int\frac{d\mathbf{q}}{(2\pi)^2} \int_{-\infty}^\infty\frac{dx}{2\pi}\mathrm{Im}\chi(\mathbf{q},x)\frac{\coth \frac{x}{2T}-\tanh\frac{\varepsilon_{p+q}}{2T}}{i\epsilon_n+x-\varepsilon_{p+q}}.
\end{equation}
The integral over $x$ is restricted by $\mathrm{Im}\chi(\mathbf{q},x)$ to the region $x\lesssim \omega_\mathrm{sf}\xi^{-2}$. For $\omega_\mathrm{sf} \xi^{-2}\ll T$, $\coth\frac{x}{2T}\approx 2T/x$, $\tanh\frac{\varepsilon_{p+q}}{2T}$ can be neglected, and Eq (\ref{eq:SE W}) is approximated as 
\begin{equation}\label{eq:app y integral}
\Sigma(\mathbf{p},i\epsilon_n)\approx
g^2T\chi_0\int\frac{d\mathbf{q}}{(2\pi)^2}\int\frac{dy}{\pi}\frac{\xi^2}{(1+\xi^2(\mathbf{q}-\mathbf{Q})^2)^2+y^2}\frac{1}{\omega_\mathrm{sf}\xi^{-2}y+i\epsilon_n-\varepsilon_{p+q}},
\end{equation}
where $y=\xi^2 x/\omega_\mathrm{sf}$. Performing the $y$-integral by closing the contour to avoid the pole from the electron propagator, 
\begin{equation}
\Sigma(\mathbf{p},i\epsilon_n)=g^2 T\int \frac{d\mathbf{q}}{(2\pi)^2}\frac{\chi_0}{\xi^{-2}+(\mathbf{q}-\mathbf{Q})^2}\frac{1}{i\epsilon_n+i\omega_\mathrm{sf}\xi^{-2}a\mathrm{sgn}\epsilon_n-\varepsilon_{p+q}},
\end{equation}
where $a=(1+\xi^{-2}(\mathbf{q}-\mathbf{Q})^2)$ is a quantity of order 1, and $i\omega_\mathrm{sf} \xi^{-2}a \mathrm{sgn}\epsilon_n$ can be neglected compared to $i\epsilon_n$. This leads to  Eq (\ref{eq:SE eq}) which was obtained in the static limit using Eq (\ref{eq:chi static}). 

For the current vertex function in the leading order perturbation theory, the use of Eq (\ref{eq:chi static}) is also justified in the same way as above; we can write an equation analogous to Eq (\ref{eq:VC eq}), and then split the two fermion Green's functions as was done in Sec \ref{conductivity}.


\section{Summary of the formulas in the mean-field theory of the spin density wave state}
\label{MF}

In this Appendix, we summarize the formulas of calculating the longitudinal and Hall conductivities in the mean-field SDW state. In calculating these quantities, the spin index $\sigma$ is irrelevant, giving an overall factor of 2, and will be neglected. The mean-field Hamiltonian is 
\begin{equation}
H_\mathrm{mf}=\sum_p\varepsilon_p c_p^\dagger c_p+\Delta\sum_p c_{p+Q}^\dagger c_p\equiv\sideset{}{^\prime}\sum_p{}\Psi_p^\dagger \hat{\mathcal{H}}_p\Psi_p, 
\end{equation}
with $\mathbf{Q}=(\pi,\pi)$, the two-component spinor $\Psi_p^\dagger=(c_p^\dagger,c_{p+Q}^\dagger)$, and $\hat{\mathcal{H}}_p=\begin{pmatrix} \varepsilon_p & \Delta\\ \Delta &\varepsilon_{p+Q}\end{pmatrix}$. The summation in the second equality is over the magnetic Brillouin zone as indicated by the prime. 

The imaginary-time ($\tilde{\tau}$) electron Green's function in the mean-field theory is defined as 
\begin{equation}
\hat{G}(\mathbf{p},\tilde{\tau})_{ab}=-<T_\tau \Psi_{p,a}(\tilde{\tau})\Psi_{p,b}^\dagger(0)>,
\end{equation}
with the corresponding retarded function 
\begin{equation}\label{eq:G MF}
\hat{G}^R(\mathbf{p},\epsilon)=\frac{1}{\epsilon-\hat{\mathcal{H}}_p+i/2\tau}=\frac{\begin{pmatrix}\epsilon-\varepsilon_{p+Q}+i/2\tau & \Delta \\ \Delta & \epsilon-\varepsilon_p+i/2\tau\end{pmatrix}}{(\epsilon-\varepsilon_p+i/2\tau)(\epsilon-\varepsilon_{p+Q}+i/2\tau)-\Delta^2},
\end{equation}
where we have introduced a finite lifetime $\tau$. 

\begin{figure}
\centering
\includegraphics[width=.8\textwidth]{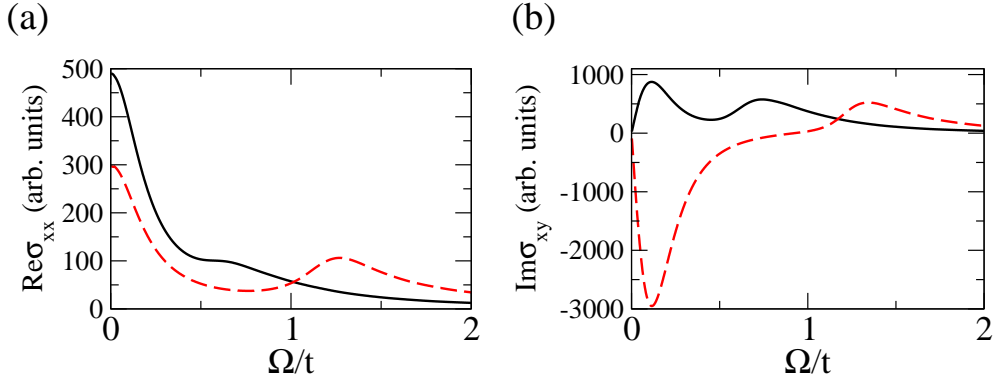}
\caption{\small{The conductivities in the mean-field theory. (a): $\mathrm{Re}\sigma_{xx}(\Omega)$ obtained from Eq (\ref{eq:conductivity MF}) with $\Delta=0.3t$ (solid line) and $\Delta=0.6t$ (dashed line). (b): $\mathrm{Im}\sigma_{xy}(\Omega)$ obtained from Eq (\ref{eq:Hall MF}) with $\Delta=0.3t$ (solid line) and $\Delta=0.6t$ (dashed line).}}
\label{fig:MF}
\end{figure}

The real part of the longitudinal conductivity $\sigma_{xx}(\Omega)$ is given by 
\begin{equation}\label{eq:conductivity MF}
\mathrm{Re}\sigma_{xx}(\Omega)=4\sigma_Q\sideset{}{^\prime}\sum_p\int_{-\infty}^\infty \frac{d\omega}{2\pi}\frac{f(\omega)-f(\omega+\Omega)}{\Omega} \mathrm{Tr}\Bigl\{\hat{v}_p^x \mathrm{Im}[\hat{G}^R(\mathbf{p},\omega)]\hat{v}_p^x \mathrm{Im}[\hat{G}^R(\mathbf{p},\omega+\Omega)]\Bigr\},
\end{equation}
where $\hat{v}_p^x=\begin{pmatrix} v_p^x & 0\\ 0 & v_{p+Q}^x \end{pmatrix}$. Fig \ref{fig:MF} (a) shows $\mathrm{Re}\sigma_{xx}(\Omega)$ in the mean-field theory for $\Delta=0.3t$ (solid line) and $0.6t$ (dashed line). 

The imaginary part of the Hall conductivity is given by\cite{Zimmers07,Voruganti92}
\begin{equation}\label{eq:Hall MF}
\mathrm{Im}\sigma_{xy}(\Omega)=\frac{\pi}{2}\sigma_Q\frac{Ba^2}{\Phi_0}\frac{1}{\Omega}\sideset{}{^\prime}\sum_p\Bigl\{\sum_{s=1}^{2}\mathcal{R}^{\mathrm{intra}(s)}_p\mathrm{Im}\Pi_R^{\mathrm{intra}(s)}(\mathbf{p},\Omega)+\sum_{s=1}^{4}\mathcal{R}^{\mathrm{inter}(s)}_p\mathrm{Im}\Pi_R^{\mathrm{inter}(s)}(\mathbf{p},\Omega)\Bigr\},
\end{equation}
where 
\begin{equation}
\begin{split}
\mathrm{Im} \Pi_R^{\mathrm{intra}(1)}(\mathbf{p},\Omega)&=
\int_{-\infty}^{\infty}\frac{d\omega}{2\pi}[f(\omega)-f(\omega+\Omega)]\\
&\qquad\qquad
\times\Bigl\{\mathrm{Re}G^+(\mathbf{p},\omega+\Omega)
A^+(\mathbf{p},\omega+\Omega)A^+(\mathbf{p},\omega)\\
&\qquad\qquad\qquad
-\mathrm{Re}G^+(\mathbf{p},\omega)A^+(\mathbf{p},\omega)A^+(\mathbf{p},\omega+\Omega)
\Bigr\},
\end{split}
\end{equation}
\begin{equation}
\begin{split}
{\rm Im} \Pi_R^{{\rm intra}(2)}(\mathbf{p},\Omega)&=
\int_{-\infty}^{\infty}\frac{d\omega}{2\pi}[f(\omega)-f(\omega+\Omega)]\\
&\qquad\qquad
\times\Bigl\{
\mathrm{Re}G^-(\mathbf{p},\omega+\Omega)
A^-(\mathbf{p},\omega+\Omega)A^-(\mathbf{p},\omega)\\
&\qquad\qquad\qquad
-\mathrm{Re}
G^-(\mathbf{p},\omega)A^-(\mathbf{p},\omega)A^-(\mathbf{p},\omega+\Omega)
\Bigr\},
\end{split}
\end{equation}
\begin{equation}
\begin{split}
\mathrm{Im} \Pi_R^{\mathrm{inter}(1)}(\mathbf{p},\Omega)&=
\int_{-\infty}^{\infty}\frac{d\omega}{2\pi}[f(\omega)-f(\omega+\Omega)]\\
&\qquad\quad
\times\Bigl\{
\mathrm{Re}G^+(\mathbf{p},\omega+\Omega)
A^+(\mathbf{p},\omega+\Omega)A^-(\mathbf{p},\omega)\\
&\qquad\qquad\qquad
-\mathrm{Re}G^+(\mathbf{p},\omega)A^+(\mathbf{p},\omega)A^-(\mathbf{p},\omega+\Omega)
\Bigr\},
\end{split}
\end{equation}
\begin{equation}
\begin{split}
\mathrm{Im} \Pi_R^{\mathrm{inter}(2)}(\mathbf{p},\Omega)&=
\int_{-\infty}^{\infty}\frac{d\omega}{2\pi}[f(\omega)-f(\omega+\Omega)]\\
&\qquad\quad
\times\Bigl\{\mathrm{Re}G^-(\mathbf{p},\omega+\Omega)
A^-(\mathbf{p},\omega+\Omega)A^+(\mathbf{p},\omega)\\
&\qquad\qquad\qquad
-\mathrm{Re}G^-(\mathbf{p},\omega)A^-(\mathbf{p},\omega)A^+(\mathbf{p},\omega+\Omega)
\Bigr\},
\end{split}
\end{equation}
\begin{equation}
\begin{split}
\mathrm{Im} \Pi_R^{\mathrm{inter}(3)}(\mathbf{p},\Omega)&=
\int_{-\infty}^{\infty}\frac{d\omega}{2\pi}[f(\omega)-f(\omega+\Omega)]/2\\
&\qquad
\times\Biggl\{\Bigl(\mathrm{Re}G^+(\mathbf{p},\omega+\Omega)A^-(\mathbf{p},\omega+\Omega)
\\&\qquad\qquad\qquad
+\mathrm{Re}G^-(\mathbf{p},\omega+\Omega)A^+(\mathbf{p},\omega+\Omega)\Bigr)
A^+(\mathbf{p},\omega)\\
&\qquad\qquad
-\Bigl(\mathrm{Re}
G^+(\mathbf{p},\omega)A^-(\mathbf{p},\omega)\\
&\qquad\qquad\qquad
+\mathrm{Re}G^-(\mathbf{p},\omega)A^+(\mathbf{p},\omega)\Bigr)
A^+(\mathbf{p},\omega+\Omega)
\Biggr\},
\end{split}
\end{equation}
\begin{equation}
\begin{split}
\mathrm{Im} \Pi_R^{\mathrm{inter}(4)}(\mathbf{p},\Omega)&=
\int_{-\infty}^{\infty}\frac{d\omega}{2\pi}[f(\omega)-f(\omega+\Omega)]/2\\
&\qquad
\times\Biggl\{
\Bigl(\mathrm{Re}G^+(\mathbf{p},\omega+\Omega)A^-(\mathbf{p},\omega+\Omega)\\
&\qquad\qquad\qquad
+\mathrm{Re}G^-(\mathbf{p},\omega+\Omega)
A^+(\mathbf{p},\omega+\Omega)\Bigr)A^-(\mathbf{p},\omega)\\
&\qquad\qquad
-\Bigl(\mathrm{Re}
G^+(\mathbf{p},\omega)A^-(\mathbf{p},\omega)\\
&\qquad\qquad\qquad
+\mathrm{Re}G^-(\mathbf{p},\omega)A^+(\mathbf{p},\omega)\Bigr)
A^-(\mathbf{p},\omega+\Omega)
\Biggr\},
\end{split}
\end{equation}
\begin{equation}
\mathcal{R}^{\mathrm{intra}(1)}_p=(E^{+y}_p)^2 E^{+xx}_p
+(E^{+x}_p)^2 E^{+yy}_p-2E^{+x}_p E^{+y}_p E^{+xy}_p,
\end{equation}
\begin{equation}
\mathcal{R}^{\mathrm{intra}(2)}_p=(E^{-y}_p)^2 E^{-xx}_p
+(E^{-x}_p)^2 E^{-yy}_p-2E^{-x}_p E^{-y}_p E^{-xy}_p,
\end{equation}
\begin{equation}
\begin{split}
\mathcal{R}^{{\rm inter}(1)}_p=&\sin^2 2\theta_p \left(
h_p^y g_p^y h_p^{xx}+h_p^x g_p^x h_p^{yy}
-h_p^x g_p^y h_p^{xy}-h_p^y g_p^x h_p^{xy}\right)\\
&+\sin^2 2\theta_p\frac{h_p}{\sqrt{h_p^2+\Delta^2}}
\left((h_p^y)^2h_p^{xx}+(h_p^x)^2 h_p^{yy}
-2h_p^x h_p^y h_p^{xy}\right)\\
&+\frac{\sin^3 2\theta_p}{\Delta}(h_p^x g_p^y-h_p^y g_p^x)^2,
\end{split}
\end{equation}
\begin{equation}
\begin{split}
\mathcal{R}^{{\rm inter}(2)}_p=&\sin^2 2\theta_p \left(
h_p^y g_p^y h_p^{xx}+h_p^x g_p^x h_p^{yy}
-h_p^x g_p^y h_p^{xy}-h_p^y g_p^x h_p^{xy}\right)\\
&-\sin^2 2\theta_p\frac{h_p}{\sqrt{h_p^2+\Delta^2}}
\left((h_p^y)^2h_p^{xx}+(h_p^x)^2 h_p^{yy}
-2h_p^x h_p^y h_p^{xy}\right)\\
&-\frac{\sin^3 2\theta_p}{\Delta}(h_p^x g_p^y-h_p^y g_p^x)^2,
\end{split}
\end{equation}
\begin{equation}
\begin{split}
\mathcal{R}^{{\rm inter}(3)}_p=&\sin^2 2\theta_p 
\left((h_p^y)^2g_p^{xx}+(h_p^x)^2 g_p^{yy}
-2h_p^x h_p^y g_p^{xy}\right)\\
&+\sin^2 2\theta_p \left(
h_p^y g_p^y h_p^{xx}+h_p^x g_p^x h_p^{yy}
-h_p^x g_p^y h_p^{xy}-h_p^y g_p^x h_p^{xy}\right)\\
&+\sin^2 2\theta_p\frac{2h_p}{\sqrt{h_p^2+\Delta^2}}
\left((h_p^y)^2h_p^{xx}+(h_p^x)^2 h_p^{yy}
-2h_p^x h_p^y h_p^{xy}\right)\\
&+2\frac{\sin^32\theta_p}{\Delta}(h_p^y g_p^x-h_p^x g_p^y)^2,
\end{split}
\end{equation}
\begin{equation}
\begin{split}
\mathcal{R}^{{\rm inter}(4)}_p=&\sin^2 2\theta_p 
\left((h_p^y)^2g_p^{xx}+(h_p^x)^2 g_p^{yy}
-2h_p^x h_p^y g_p^{xy}\right)\\
&+\sin^2 2\theta_p \left(
h_p^y g_p^y h_p^{xx}+h_p^x g_p^x h_p^{yy}
-h_p^x g_p^y h_p^{xy}-h_p^y g_p^x h_p^{xy}\right)\\
&-\sin^2 2\theta_p\frac{2h_p}{\sqrt{h_p^2+\Delta^2}}
\left((h_p^y)^2h_p^{xx}+(h_p^x)^2 h_p^{yy}
-2h_p^x h_p^y h_p^{xy}\right)\\
&-2\frac{\sin^32\theta_p}{\Delta}(h_p^y g_p^x-h_p^x g_p^y)^2.
\end{split}
\end{equation}
In the above, we have introduced notations $G^{\pm}(\mathbf{p},\epsilon)=1/(\epsilon-E_p^{\pm}+i/2\tau)$ with $E_p^\pm=g_p\pm\sqrt{h_p^2+\Delta^2}$, $g_p=(\varepsilon_p+\varepsilon_{p+Q})/2$, $h_p=(\varepsilon_p-\varepsilon_{p+Q})/2$, $\tan\theta_p=(h_p-\sqrt{h_p^2+\Delta^2})/\Delta$, $A^\pm(\mathbf{p},\omega)=-2\mathrm{Im}G^{\pm}(\mathbf{p},\omega)$, and have used the short-hand notation in which the superscript $x,y$ on the energy functions denotes derivative with respect to the corresponding momentum, \textit{e.g.}, $E_p^{+x}=\partial E_p^+/\partial p_x$, $E_p^{+xx}=\partial^2 E_p^+/\partial p_x^2$, $g_p^x=\partial g_p/\partial p_x$, $g^{xy}_p=\partial^2g_p/\partial p_x\partial p_y$, $\cdots$. Figure \ref{fig:MF} (b) shows $\mathrm{Im}\sigma_{xy}(\Omega)$ in the mean-field theory for $\Delta=0.3t$ (solid line) and $\Delta=0.6t$ (dashed line).


\section{Explicit expression for $\mathrm{Im}\sigma_{xy}$ in Sec \ref{Hall}}
\label{sxy}

We now present the explicit expression for $\mathrm{Im}\sigma_{xy}(\Omega)$ in Sec \ref{Hall},
\begin{equation}\label{eq:Hall}
\mathrm{Im}\sigma_{xy}(\Omega)=\frac{\pi}{2}\sigma_Q\frac{Ba^2}{\Phi_0}\frac{1}{\Omega}\int \frac{d\mathbf{p}}{(2\pi)^2}\mathrm{Im}\bigl\{S^R_1-S^R_2+S^R_3+S^R_4-S^R_5\bigr\},
\end{equation}
where 
\begin{equation}
\mathrm{Im}S^R_i(\mathbf{p},\omega)=\int_{-\infty}^\infty\frac{d\epsilon}{2\pi}\bigl[f(\epsilon)-f(\epsilon+\omega)\bigr]\mathrm{Re}\bigl\{U_{i1}-U_{i2}\bigr\},
\end{equation}
with 
\begin{multline}
U_{11}=\bigl[\partial_y\Gamma_x(\mathbf{p},\epsilon^+,\epsilon+\omega^+)v_p^y-\partial_y\Gamma_y(\mathbf{p},\epsilon^+,\epsilon+\omega^+)v_p^x\bigr]\\
\times\bigl[G^R(\mathbf{p},\epsilon)\partial_xG^R(\mathbf{p},\epsilon+\omega)-G^R(\mathbf{p},\epsilon+\omega)\partial_xG^R(\mathbf{p},\epsilon)\bigr],
\end{multline}
\begin{multline}
U_{12}=\bigl[\partial_y\Gamma_x(\mathbf{p},\epsilon^-,\epsilon+\omega^+)v_p^y-\partial_y\Gamma_y(\mathbf{p},\epsilon^-,\epsilon+\omega^+)v_p^x\bigr]\\
\times\bigl[G^A(\mathbf{p},\epsilon)\partial_xG^R(\mathbf{p},\epsilon+\omega)-G^R(\mathbf{p},\epsilon+\omega)\partial_xG^A(\mathbf{p},\epsilon)\bigr],
\end{multline}
\begin{multline}
U_{21}=\bigl[\partial_x\Gamma_x(\mathbf{p},\epsilon^+,\epsilon+\omega^+)v_p^y-\partial_x\Gamma_y(\mathbf{p},\epsilon^+,\epsilon+\omega^+)v_p^x\bigr]\\
\times\bigl[G^R(\mathbf{p},\epsilon)\partial_yG^R(\mathbf{p},\epsilon+\omega)-G^R(\mathbf{p},\epsilon+\omega)\partial_yG^R(\mathbf{p},\epsilon)\bigr],
\end{multline}
\begin{multline}
U_{22}=\bigl[\partial_x\Gamma_x(\mathbf{p},\epsilon^-,\epsilon+\omega^+)v_p^y-\partial_x\Gamma_y(\mathbf{p},\epsilon^-,\epsilon+\omega^+)v_p^x\bigr]\\
\times\bigl[G^A(\mathbf{p},\epsilon)\partial_yG^R(\mathbf{p},\epsilon+\omega)-G^R(\mathbf{p},\epsilon+\omega)\partial_yG^A(\mathbf{p},\epsilon)\bigr],
\end{multline}
\begin{multline}
U_{31}=\bigl[\Gamma_x(\mathbf{p},\epsilon^+,\epsilon+\omega^+)v_p^y-\Gamma_y(\mathbf{p},\epsilon^+,\epsilon+\omega^+)v_p^x\bigr]\\
\times\bigl[\partial_xG^R(\mathbf{p},\epsilon+\omega)\partial_yG^R(\mathbf{p},\epsilon)-\partial_xG^R(\mathbf{p},\epsilon)\partial_yG^R(\mathbf{p},\epsilon+\omega)\bigr],
\end{multline}
\begin{multline}
U_{32}=\bigl[\Gamma_x(\mathbf{p},\epsilon^-,\epsilon+\omega^+)v_p^y-\Gamma_y(\mathbf{p},\epsilon^-,\epsilon+\omega^+)v_p^x\bigr]\\
\times\bigl[\partial_xG^R(\mathbf{p},\epsilon+\omega)\partial_yG^A(\mathbf{p},\epsilon)-\partial_xG^A(\mathbf{p},\epsilon)\partial_yG^R(\mathbf{p},\epsilon+\omega)\bigr],
\end{multline}
\begin{equation}
U_{41}=v_p^y(\varepsilon_{p+Q}^{xx}v_{p+Q}^y-\varepsilon_{p+Q}^{xy}v_{p+Q}^x)G^R(\mathbf{p},\epsilon)G^R(\mathbf{p},\epsilon+\omega)I(\mathbf{p},\epsilon^+,\epsilon+\omega^+),
\end{equation}
\begin{equation}
U_{42}=v_p^y(\varepsilon_{p+Q}^{xx}v_{p+Q}^y-\varepsilon_{p+Q}^{xy}v_{p+Q}^x)G^A(\mathbf{p},\epsilon)G^R(\mathbf{p},\epsilon+\omega)I(\mathbf{p},\epsilon^-,\epsilon+\omega^+),
\end{equation}
\begin{equation}
U_{51}=v_p^x(\varepsilon_{p+Q}^{xy}v_{p+Q}^y-\varepsilon_{p+Q}^{yy}v_{p+Q}^x)G^R(\mathbf{p},\epsilon)G^R(\mathbf{p},\epsilon+\omega)I(\mathbf{p},\epsilon^+,\epsilon+\omega^+),
\end{equation}
\begin{equation}
U_{52}=v_p^x(\varepsilon_{p+Q}^{xy}v_{p+Q}^y-\varepsilon_{p+Q}^{yy}v_{p+Q}^x)G^A(\mathbf{p},\epsilon)G^R(\mathbf{p},\epsilon+\omega)I(\mathbf{p},\epsilon^-,\epsilon+\omega^+).
\end{equation}
In writing these equations, we have used the notations $\epsilon^{\pm}=\epsilon\pm i\delta$, and $\epsilon+\omega^+=\epsilon+\omega+i\delta$. The analytically continued vertex functions are 
\begin{equation}
\Gamma_\alpha(\mathbf{p},\epsilon^+,\epsilon+\omega^+)=v_p^\alpha+\frac{v_{p+Q}^\alpha}{\omega}\bigl(\Sigma^R(\mathbf{p},\epsilon)-\Sigma^R(\mathbf{p},\epsilon+\omega)\bigr),
\end{equation}
and
\begin{equation}
\Gamma_\alpha(\mathbf{p},\epsilon^-,\epsilon+\omega^+)=v_p^\alpha+\frac{v_{p+Q}^\alpha}{\omega+i/\tau}\bigl(\Sigma^A(\mathbf{p},\epsilon)-\Sigma^R(\mathbf{p},\epsilon+\omega)\bigr).
\end{equation}
The functions $I(\mathbf{p},\epsilon^{\pm},\epsilon+\omega^+)$ are defined as 
\begin{equation}
I(\mathbf{p},\epsilon^+,\epsilon+\omega^+)=-\frac{\gamma^R(\mathbf{p},\epsilon)+\gamma^R(\mathbf{p},\epsilon+\omega)}{\omega}-2\frac{\Sigma^R(\mathbf{p},\epsilon)-\Sigma^R(\mathbf{p},\epsilon+\omega)}{\omega^2},
\end{equation}
and 
\begin{equation}
I(\mathbf{p},\epsilon^-,\epsilon+\omega^+)=-\frac{\gamma^A(\mathbf{p},\epsilon)+\gamma^R(\mathbf{p},\epsilon+\omega)}{\omega+i/\tau}-2\frac{\Sigma^A(\mathbf{p},\epsilon)-\Sigma^R(\mathbf{p},\epsilon+\omega)}{(\omega+i/\tau)^2},
\end{equation}
where $\gamma^{A,R}(\mathbf{p},\epsilon)=\partial\Sigma^{A,R}(\mathbf{p},\epsilon)/\partial\epsilon$. The real part of $\sigma_{xy}(\Omega)$ is obtained via the Kramers-Kr\"onig relation, Eq (\ref{eq:KK}), with $\Sigma_{if}^R$ replaced by $\sigma_{xy}$.

\bibliography{LRA_submit}

\end{document}